\documentclass[fleqn,10pt]{wlscirep}
\usepackage[utf8]{inputenc}
\usepackage[T1]{fontenc}

\usepackage{fix-cm}
\usepackage[normalem]{ulem}
\usepackage{graphicx}%
\usepackage{multirow}%
\usepackage{amsmath,amssymb,amsfonts}%
\usepackage{amsthm}%
\usepackage{mathrsfs}%
\usepackage[title]{appendix}%
\usepackage{textcomp}%
\usepackage{manyfoot}%
\usepackage{booktabs}%
\usepackage{algorithm}%
\usepackage{algorithmicx}%
\usepackage{algpseudocode}%
\usepackage{listings}%
\usepackage{textcomp}%
\usepackage{float}
\usepackage{comment}
\usepackage{xurl}

\title{Continental-scale assessment of spatial food market accessibility in Africa using open geospatial data}

\author[1, 2 *]{Robert Benassai-Dalmau}

\author[3 *]{Vasiliki Voukelatou}

\author[4,1]{Rossano Schifanella}
\author[1,5]{Stefania Fiandrino}

\author[1]{Daniela Paolotti}

\author[1 *]{Kyriaki Kalimeri}

\affil[1]{ISI Foundation, Via della Rocca, 20, Turin, 10123, Italy}

\affil[2]{Internet Interdisciplinary Institute, Universitat Oberta de Catalunya, Rambla del Poblenou, 156, Barcelona, 08018, Spain}

\affil[3]{Forecasting and Early Warning Unit, United Nations World Food Programme, Via Cesare Giulio Viola 68, Rome, 00148, Italy}

\affil[4]{University of Turin, Computer Science Department, Turin, Italy}

\affil[5]{Department of Computer, Control, and Management Engineering Antonio Ruberti, Sapienza University of Rome, V. Ariosto, 25, Rome, 00185,
Italy}

\affil[*]{Corresponding Authors: rbenassai@uoc.edu, vasiliki.voukelatou@wfp.org, kyriaki.kalimeri@isi.it}


\begin{abstract}

Food market accessibility is a critical yet underexplored dimension of food systems, particularly in low- and middle-income countries. In this paper, we present a continent-wide assessment of spatial food market accessibility in Africa, integrating open geospatial data from OpenStreetMap and the World Food Programme. We compare three complementary metrics: travel time to the nearest market, market availability within a 30-minute threshold, and an entropy-based measure of spatial distribution, to quantify accessibility across diverse settings. We find pronounced disparities in accessibility: rural and economically disadvantaged populations face substantially longer travel times and reduced market availability, with some areas requiring several hours of travel. These accessibility patterns align with socioeconomic stratification, as measured by the Relative Wealth Index, and moderately correlate with food insecurity levels, assessed using the Integrated Food Security Phase Classification. Overall, results suggest that access to food markets reflects broader geographic and economic inequalities and plays a relevant role in shaping food security outcomes. Despite limitations related to incomplete and spatially heterogeneous market data coverage, this framework provides a scalable, data-driven approach for identifying relative structural market accessibility gaps, supporting equitable infrastructure planning and spatially informed food security analyses across diverse African contexts.
\end{abstract}
\begin{document}

\flushbottom
\maketitle
%
%
\thispagestyle{empty}

\section*{Introduction}\label{sec1}

Access to healthy and affordable food is a fundamental human right, included in \textit{Article 25} of the Universal Declaration of Human Rights \cite{nations1948universal}. 
According to the World Food Programme (WFP) \cite{TheWorldFoodProgramme2024}, while global food production is sufficient to feed the world’s population, disparities in food availability and accessibility persist, particularly across the African continent.
Further, the disparity between rural and urban African regions is pronounced, with rural areas often characterized by limited infrastructure, poverty, and constrained market access \cite{bonuedi2020enabling}. 
Addressing these disparities in food access is a global priority, as reflected in the United Nations Sustainable Development Goals (SDGs), where ending hunger is the second target for 2030.

In high-income countries, the concept of ``food deserts'', areas with limited access to affordable and nutritious food, is well documented \cite{lang1998access}, and a variety of quantitative methods have been used to map and analyze these areas \cite{Raja2008, Larsen2008, Short2007}. In contrast, evidence from low- and middle-income countries (LMICs), especially in Africa, remains sparse, fragmented, and often constrained by limited data coverage.  Field-based household surveys, while valuable, are resource-intensive and geographically restricted, making it challenging to conduct standardized, scalable assessments of food market access \cite{Usman2022,Chi2022, Fraval2019,Abay2017,ahmed2017status, Stifel2017, hoddinott2015cows}.

Among other factors, market proximity has been shown to influence dietary diversity and household reliance on subsistence agriculture in African contexts \cite{nandi2021,hirvonen2017agricultural}. 
Given the intensive resources required to conduct repeated field surveys, harnessing openly available, continent-wide spatial food-market datasets presents a promising way to expand and standardize analyses of market accessibility.
Since the 1970s, accessibility to primary services has been a central concern in policy-making and social science research, aimed at providing quantitative and scalable assessments for policymakers and practitioners \cite{wachs1973physical, morris1979accessibility, levine2020century}. The concept of accessibility is highly multidimensional, encompassing, for instance, availability and affordability \cite{penchansky1981concept}, and considerable effort has been devoted to delineating distinct forms of access in order to capture its inherently nuanced nature. More formally, definitions of accessibility range from purely spatial, using only the shortest travel times or distance, to more complex mathematical formulations considering supply and demand ratios, mobility, and social behaviours \cite{Geurs2004, Handy1997, cum_vs_grav}.
Healthcare and education have been the primary topics of focus in accessibility research, with studies incorporating both quantitative and qualitative analyses \cite{Pan2015, Weiss2020, Palk2020, Guagliardo2004, Maina2019, rekha2020spatial, gao2016imbalance}. Although primary service accessibility is well-established in policy-making and social sciences, a generalizable, large-scale analysis of food market accessibility is still lacking, particularly in Africa, where studies remain primarily qualitative.

In this article, we explicitly focus on the spatial dimension of food market accessibility in Africa, given that detailed market-level information (e.g., product assortment, food types, or prices) is generally unavailable across most countries on the continent.

We define a food market as a small commercial area containing multiple food shops, e.g., supermarkets, bakeries, butchers and etc. \cite{WFP_MFI} (see \hyperref[subsec: kmeans]{Methods})\, and we integrate open geospatial data from OpenStreetMap (OSM) and WFP to construct a harmonized, continent-scale dataset. This validated, scalable and entirely data-driven approach provides a consistent, cross-country analysis while filling a critical research and operational gap. We focus on three key dimensions of spatial accessibility, namely, closeness, measured by travel time to the nearest market; availability, quantified by the number of reachable markets; and spatial distribution, evaluated using the entropy of a modeled travel behaviour.

We assess the consistency and correlation among these three metrics in data-restricted countries and the impact of diverse accessibility definitions on the final insights.
We shed light on market accessibility differences between urban and rural areas, as well as the areas of diverse socioeconomic status. Finally, we explore the relation between spatial accessibility to food markets and food security in Africa.

By providing a comprehensive, continent-wide assessment of market accessibility, our study offers a structured framework for identifying regions facing the most severe spatial constraints. The methodology is scalable and can support infrastructure planning, vulnerability assessments, and spatially informed food security analyses.

\begin{figure}[hbt!]
    \centering
    \includegraphics[width=1\linewidth]{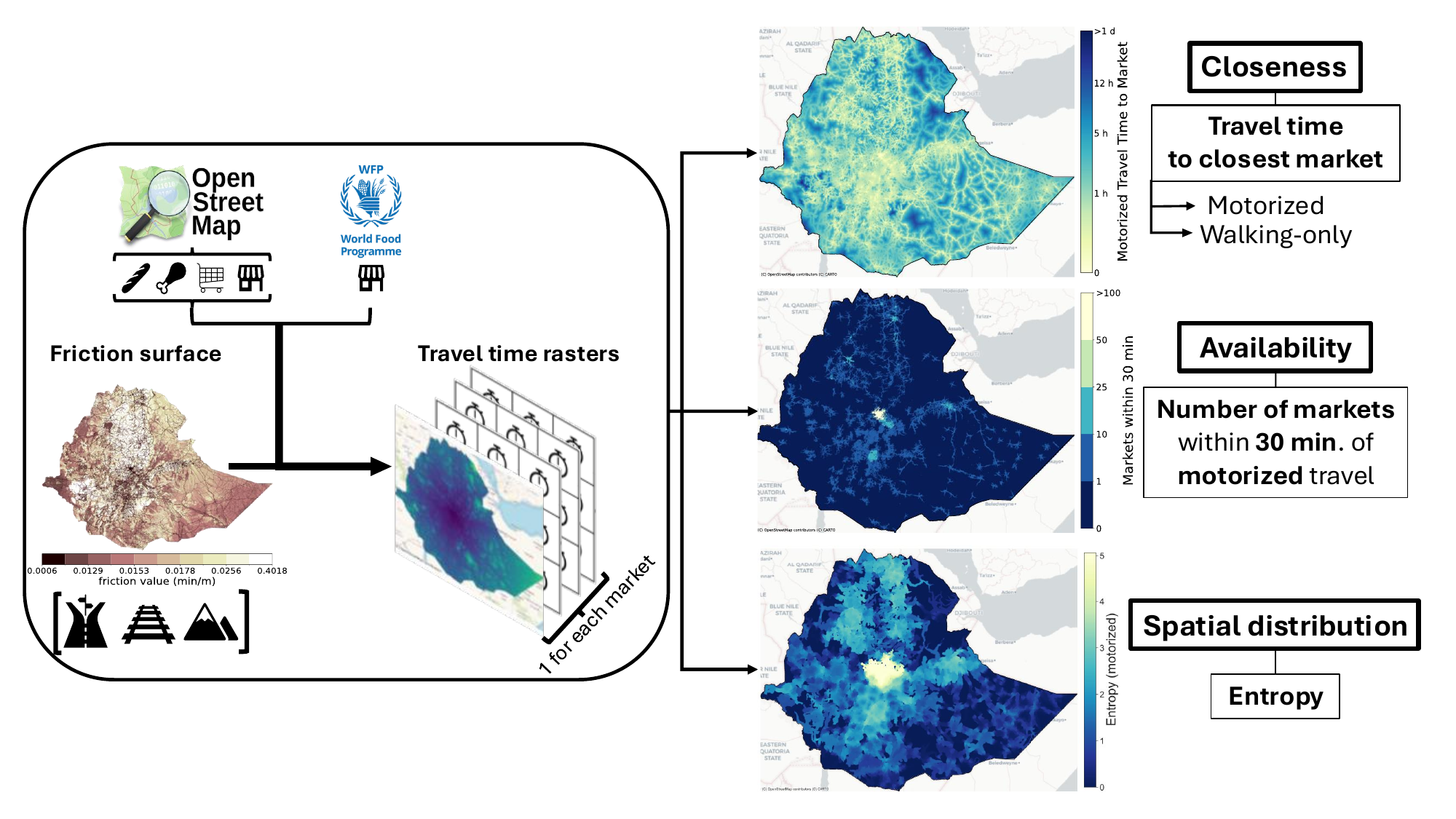}
    \caption{\textbf{Pipeline for the computation of the accessibility metrics.} This figure illustrates an example from Ethiopia. First, food shops and markets are extracted from OSM and WFP's market price and Market Functionality Index data. Market centroids are computed to harmonize both datasets. Using either a walking-only or motorized friction surface, we apply a shortest-path routing algorithm to generate a travel time raster for each market. These rasters are then used to derive accessibility metrics: travel time to the nearest market, the number of markets within a given travel time threshold, and entropy.}
    \label{fig:Pipeline}
\end{figure}

\section*{Results}

To quantify spatial market accessibility across Africa, we used a set of geospatial metrics validated in the literature \cite{Weiss2018, hansen1959accessibility, Erlander1977} and based on open data from OSM and WFP. Harmonisation of market location data from the two sources was essential to ensure comprehensive spatial coverage, reduce redundancies, and mitigate inconsistencies across sources, enabling a unified and scalable assessment of market accessibility. For this purpose, we applied an iterative \textit{k-means} clustering algorithm (see \hyperref[subsec: kmeans]{Methods}) to define market centroids. We computed travel times across a high-resolution (30 arc-second) friction surface representing both motorized and walking-only modes of transportation. Using these travel-time rasters, we utilized three accessibility metrics: (i) travel time to the closest market (closeness), (ii) the number of markets reachable within a 30-minute motorized travel window (availability) and (iii) an entropy-based index developed to capture the evenness of market distribution around each location (spatial distribution). In our context, a higher entropy implies more homogeneous market spatial distribution and vice-versa. These metrics were computed for every 30 arc-second pixel (approximately 1 $km$ at the equator) across the continent. Figure \ref{fig:Pipeline} presents an overview of the proposed framework. For the subsequent analyses concerning the relationship between market access and rurality, relative wealth and food security, we aggregated the metrics at different administrative levels, weighing the pixels by population estimates from WorldPop \cite{WorldPop_UNadj_Unconstrained_2000_2020} to account for demographic distribution (for more details, see \hyperref[subsec:WPP]{Methods}).

\subsection*{Maps of spatial market accessibility}\label{sec:results_maps_corr}

\begin{figure}[ht!]
    \centering
    \begin{tabular}{ll}
        {\bf a} & {\bf b}\\
        
        \includegraphics[trim={0cm 0 2cm 0}, clip, width = 0.49\linewidth]{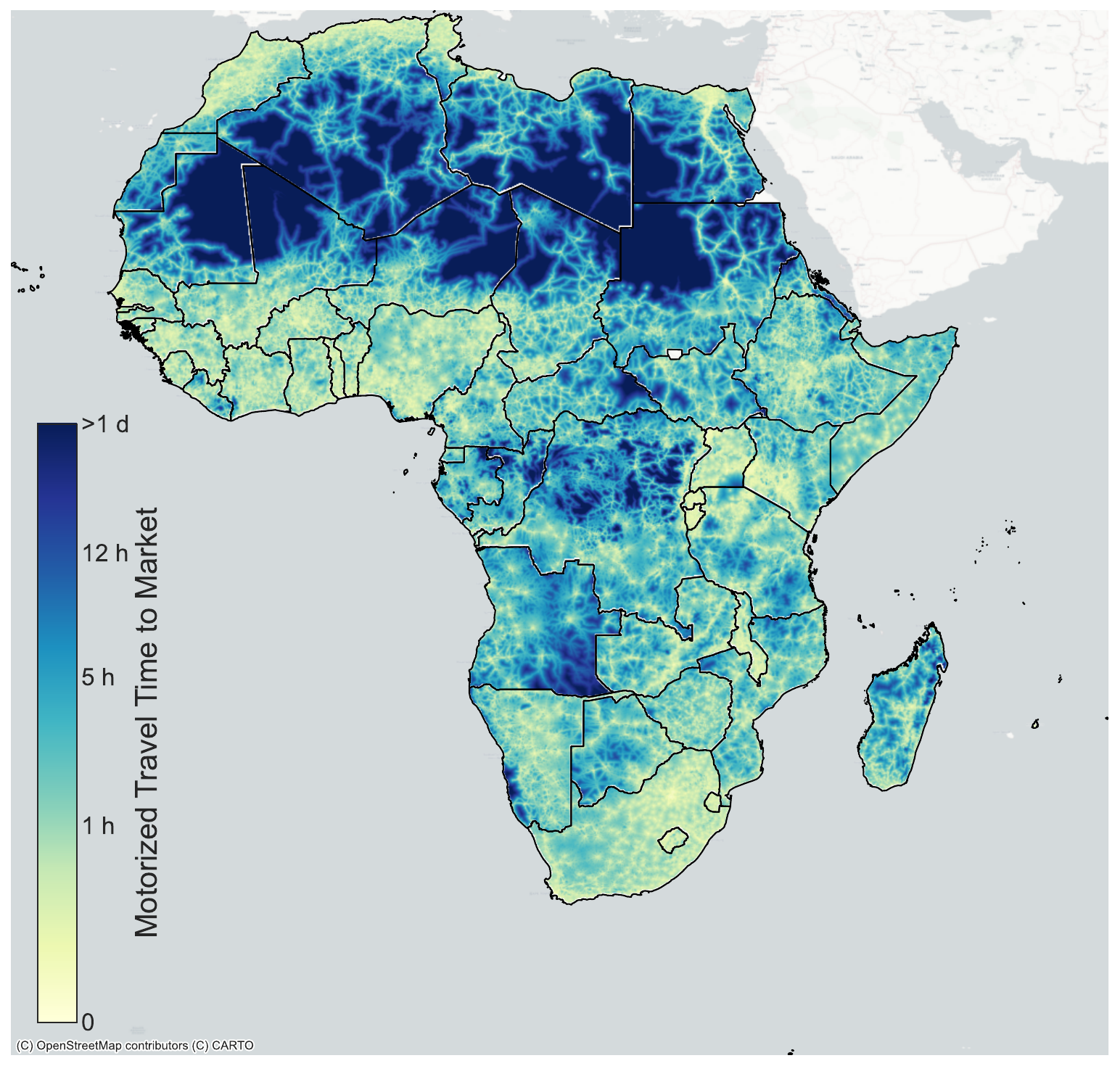}
        &
        \includegraphics[trim={0cm 0 2cm 0}, clip, width = 0.49\linewidth]{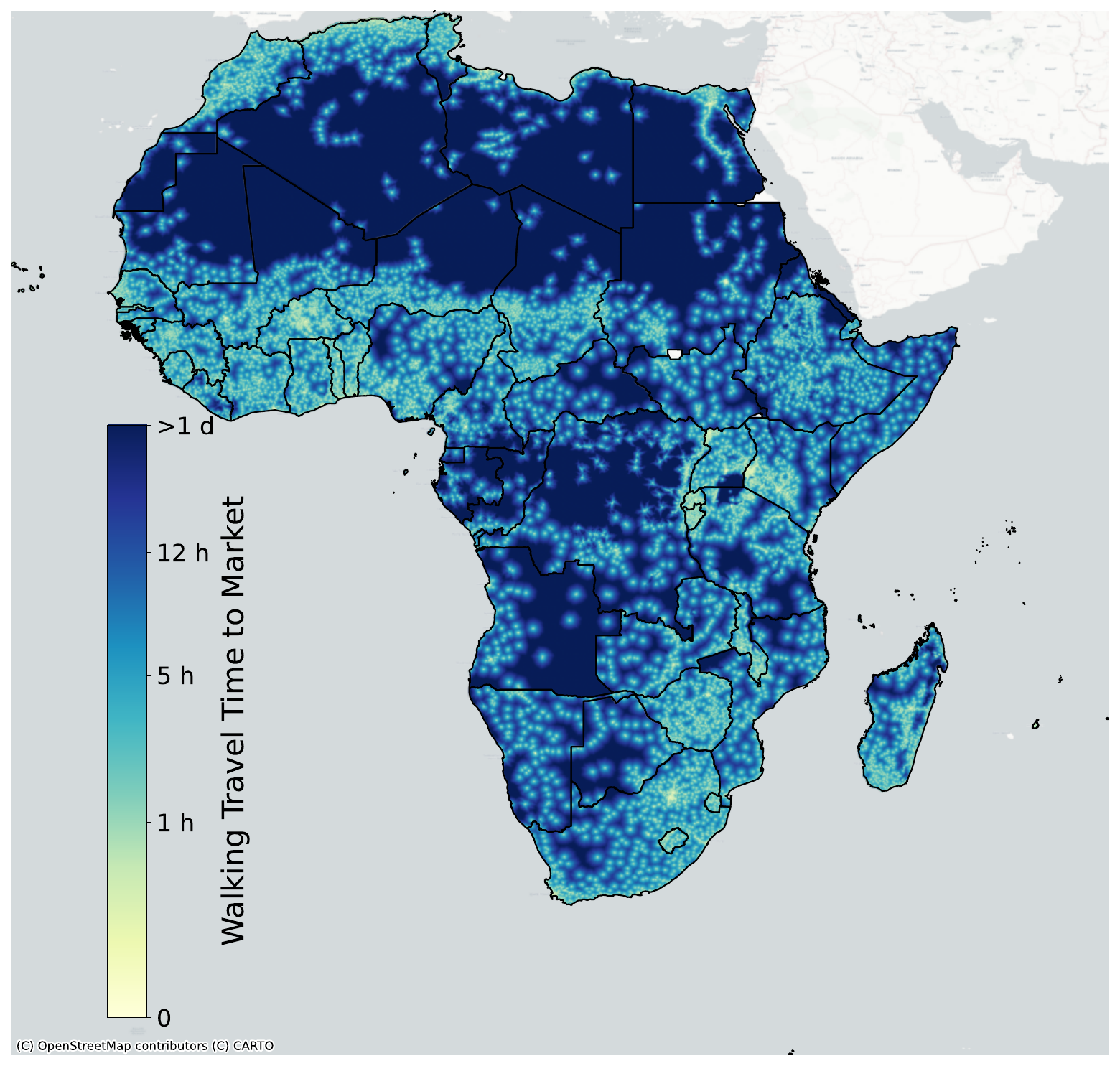}\\
        
        {\bf c} & {\bf d}\\

        \includegraphics[trim={0cm 0 2cm 0}, clip, width = 0.49\linewidth]{ 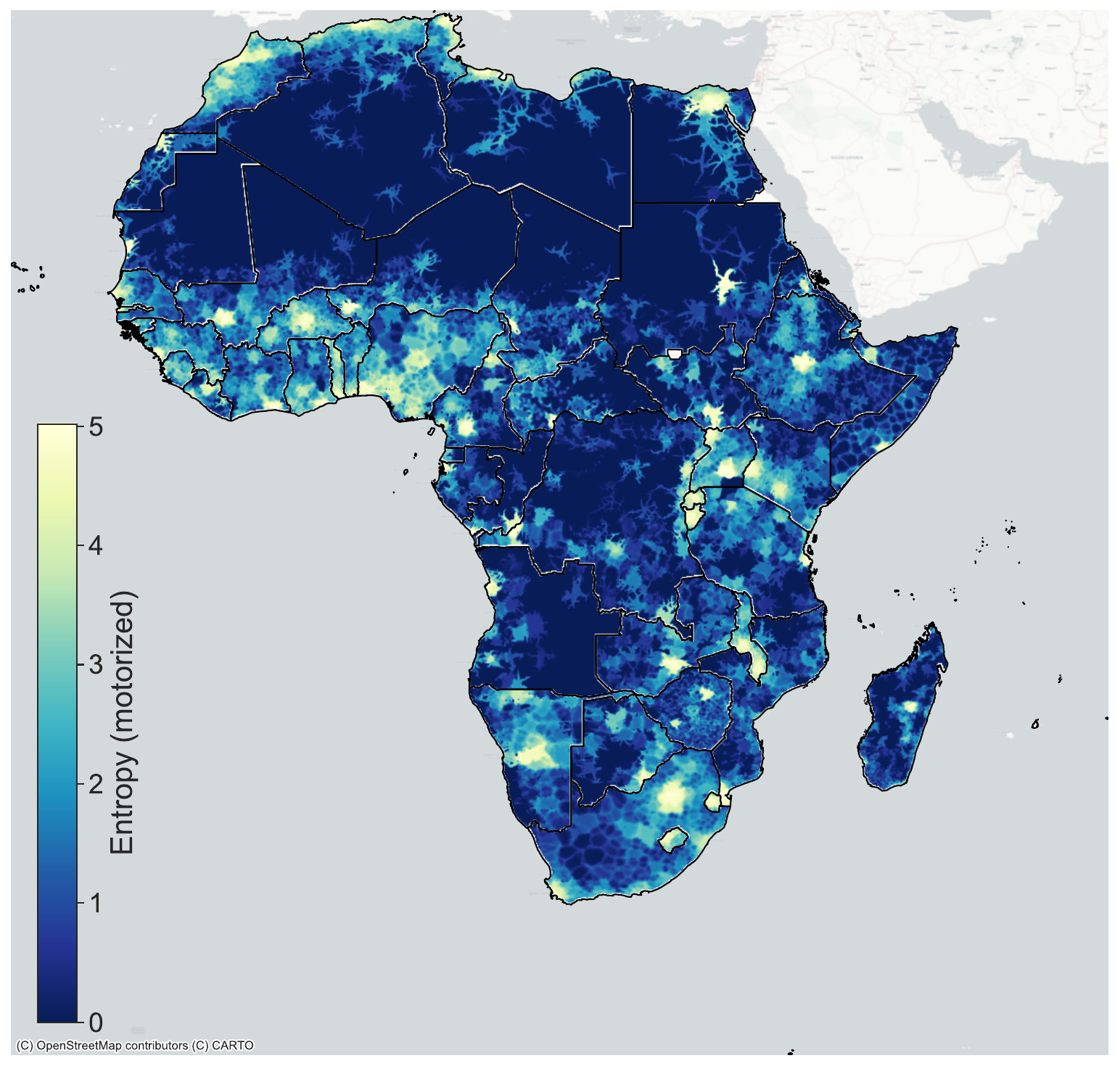}
        
        &
        \includegraphics[trim={0cm 0 2cm 0}, clip, width = 0.49\linewidth]{ 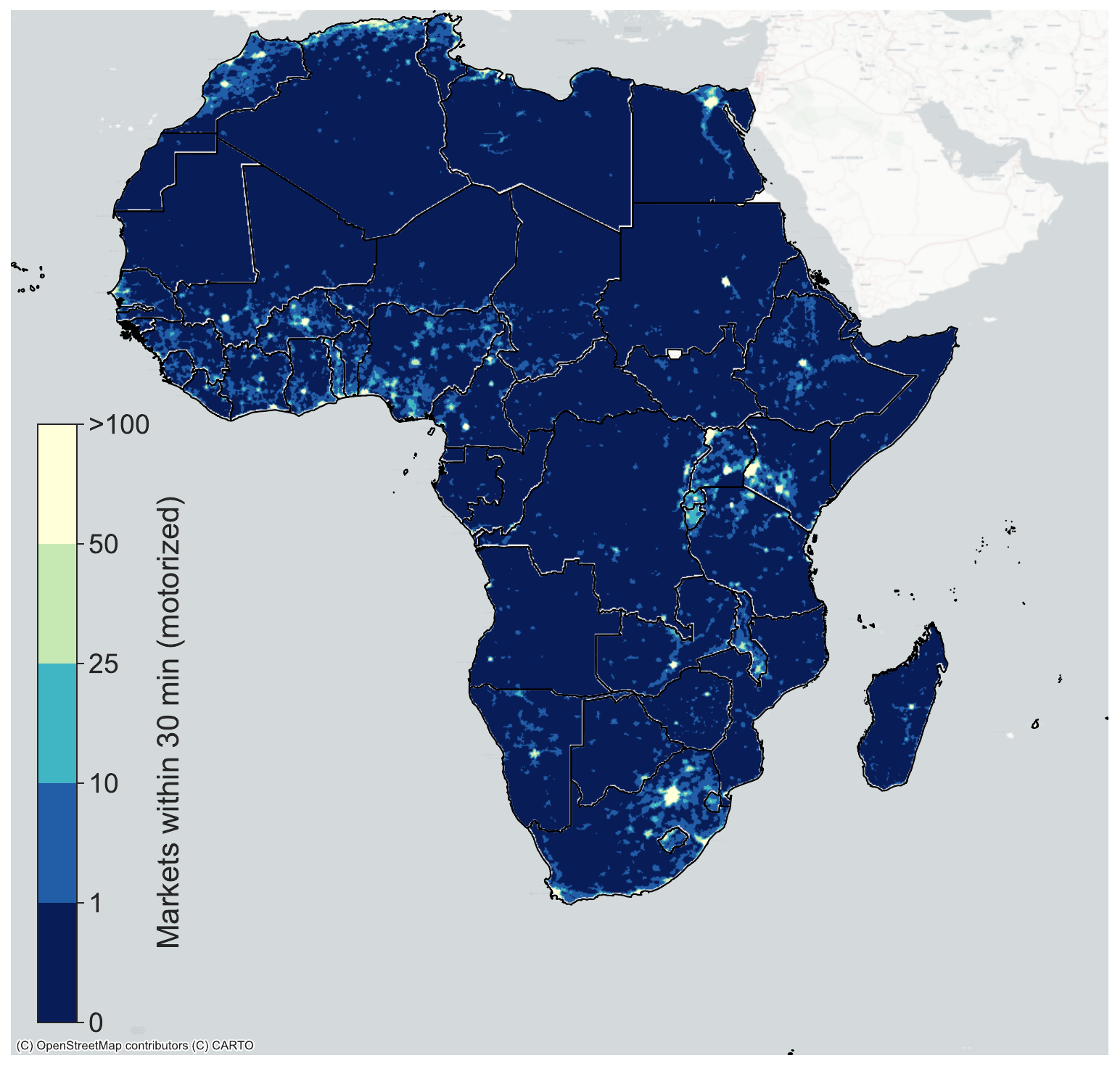}

    \end{tabular}
    \caption{\textbf{Spatial accessibility to food markets in Africa.}
    \textbf{(a)}  Motorized and \textbf{(b)} walking travel time to the closest market show widespread disparities in accessibility across the continent, with rural and remote areas facing travel times of several hours or more.\textbf{(c)} Entropy illustrates the evenness of market distribution; higher entropy suggests a more balanced spread of accessible markets, typically concentrated in capital cities and urban centers.\textbf{(d)} Availability (cumulative accessibility) reflects the number of markets reachable within 30 minutes of motorized travel, favoring urbanized regions. In the Supplementary Information section 1 Fig. 1, we also present the Entropy and Cumulative accessibility computed using walking-only travel modes.}
    \label{fig:africa_maps} 
\end{figure}

Figure \ref{fig:africa_maps} shows the maps of the metrics described above for the whole African continent. We observe similar spatial patterns in the four maps, apparently resembling the population distribution, with low accessibility in sparsely populated areas such as the Saharan desert, and better market access in more densely populated areas across all metrics. 

We examine the differences across the three metrics of market accessibility used. Analyzing the rank correlations of their population-weighted medians at the country level (see correlation matrix in the Supplementary Information section 2 Fig. 2 \textbf{a}), our findings reveal strong correlations among all spatial accessibility metrics ($|R| > 0.8$). 
Assessing the country rankings at a 1 $km^2$ pixel-level resolution, we reaffirm the concordance among the three measures (more information in the Supplementary Information section 2 Fig. 2 \textbf{b}).
This finding aligns with the patterns observed in Fig. \ref{fig:africa_maps}, demonstrating the consistency between the metrics. 
Although these accessibility definitions capture different aspects of spatial market access, their continental-scale patterns broadly align, with large population centres appearing to have the best market access. Capital cities, which are, in general, more populated, exhibit shorter travel times, higher availability (cumulative accessibility), and higher entropy (more even market spatial distribution). As an illustrative example, the five most populous cities on the continent at the time of publication according to World Population Review \cite{WPR_AfricaCities_2026} are Cairo (23.07 M), Kinshasa (17.78 M), Lagos (17.15 M), Luanda (10.03 M), and Dar es Salaam (8.56 M), located in Egypt, the Democratic Republic of the Congo, Nigeria, Angola, and Tanzania respectively. These cities demonstrate consistently high levels of market accessibility. On average, each city has at least one market reachable within 3 minutes by motorized transport or within 40 minutes on foot. Furthermore, they each host more than 90 markets accessible within 30 minutes by motorized travel and exhibit entropy values greater than 4.5. Collectively, these values place all five cities within the highest-access range in Fig. \ref{fig:africa_maps}.




\subsection*{Rural-urban disparities}

Diving deeper into the spatial inequalities between urban and rural population in Africa, we overlay our accessibility maps with population estimates from Worldpop \cite{WorldPop_UNadj_Unconstrained_2000_2020}and a four-tier rural–urban classification, namely, urban, suburban, rural, hinterland \cite{Cattaneo2020, cattaneo2021global, cattaneo2024worldwide} (see \hyperref[subsec:RWI]{Methods} for more details).

\begin{table}[ht!]
\caption{\textbf{Average accessibility for different rural-urban classifications and UN Subregions}. We show the average accessibility for urban, suburban, rural, and hinterland areas across the five UN Africa subregions: Northern, Western, Middle, Eastern, and Southern Africa (NA, WA, MA, EA, and SA, respectively). The metrics include motorized and walking travel times (in minutes) to the closest market, availability (number of markets accessible within 30 minutes of motorized travel time), and entropy, measuring the spatial distribution of markets.}
\label{tab:table_averages}
\centering
\begin{tabular}{@{}cccccc@{}}
\hline
\textbf{Accessibility} & \textbf{Region} & \textbf{Urban} & \textbf{Suburban} & \textbf{Rural} & \textbf{Hinterland} \\
\hline
\multirow{5}{*}{\begin{tabular}[c]{@{}c@{}}Motorized\\Travel Time (min)\end{tabular}}
& NA & 3.4 & 16.6 & 81.4 & 392.1 \\
& WA & 5.9 & 26.5 & 57.0 & 181.2 \\
& MA & 3.5 & 32.2 & 86.8 & 275.8 \\
& EA & 4.0 & 22.1 & 60.7 & 203.5 \\
& SA & 2.5 & 20.5 & 45.6 & 129.7 \\
\hline
\multirow{5}{*}{\begin{tabular}[c]{@{}c@{}}Walking\\Travel Time (min)\end{tabular}}
& NA & 40.5 & 149.9 & 444.1 & 951.8 \\
& WA & 74.1 & 242.0 & 297.3 & 468.7 \\
& MA & 28.9 & 249.5 & 554.2 & 1060 \\
& EA & 39.4 & 177.4 & 325.7 & 577.2 \\
& SA & 29.7 & 163.3 & 284.2 & 490.9 \\
\hline
\multirow{5}{*}{\begin{tabular}[c]{@{}c@{}}Number of markets\\within 30 minutes (motorized)\\(Availability)\end{tabular}}
& NA & 121.1 & 31.7 & 1.3 & 4.3 \\
& WA & 71.60 & 9.20 & 0.7 & 4.6 \\
& MA & 110.6 & 13.1 & 0.2 & 0.7 \\
& EA & 154.1 & 17.7 & 1.7 & 1.0 \\
& SA & 209.6 & 23.5 & 1.7 & 1.8 \\
\hline
\multirow{5}{*}{Entropy (motorized)}
& NA & 4.26 & 3.53 & 1.92 & 1.44 \\
& WA & 3.75 & 2.70 & 1.90 & 1.56 \\
& MA & 4.10 & 2.17 & 1.29 & 0.98 \\
& EA & 4.07 & 2.76 & 1.86 & 1.16 \\
& SA & 4.69 & 2.65 & 1.70 & 1.06 \\
\hline
\end{tabular}
\end{table}

\begin{figure}[ht!]
    \centering

    \includegraphics[angle=90, width = \linewidth]{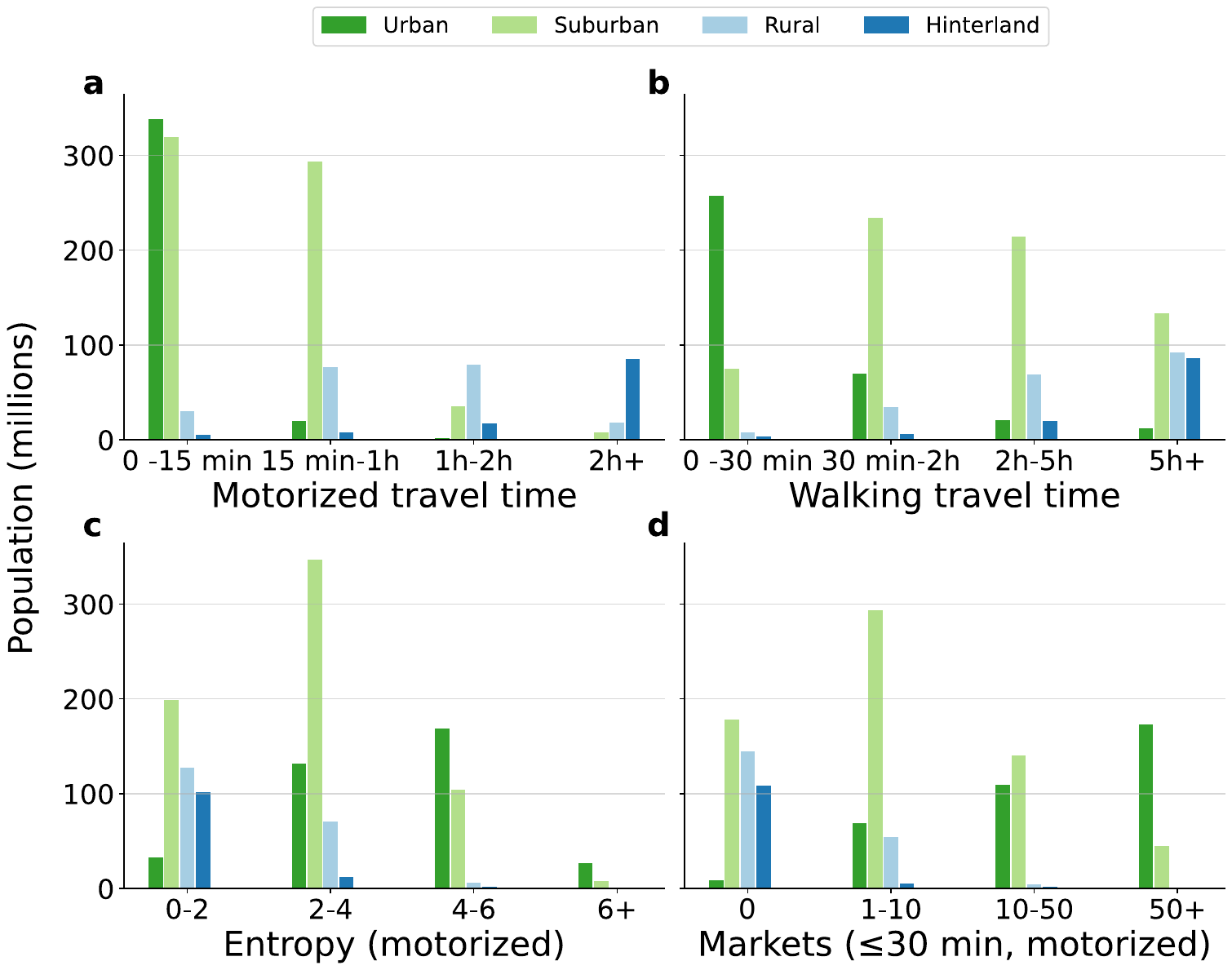}
    \caption{\textbf{Population distribution by rural-urban class and market accessibility metrics.}
    This figure shows how Africa's population is distributed across key accessibility ranges for each metric: \textbf{(a)} motorized and \textbf{(b)} walking travel time, \textbf{(c)} number of markets within 30 minutes (motorized), and \textbf{(d)} entropy. Urban populations ($\sim$360 million) overwhelmingly have better access, while rural and hinterland areas, comprising hundreds of millions, are overrepresented in categories with high travel times, low availability, and poor spatial market distribution. The bars are color-coded, expressing the rural-urban classification of the population (urban, suburban, rural, hinterland). Lower bounds are inclusive and upper bounds are exclusive for all ranges (i.e., $[L,U)$).}
    \label{fig:RurUrb_distr} 
\end{figure}

Urban residents, totaling approximately 360 million, overwhelmingly benefit from short travel times; nearly all can reach a market within 30 minutes by motorized means, while they can choose among more than 50 markets. In contrast,  approximately 250 million rural and hinterland residents lack such choices, struggling to reach even a single market within 30 minutes of motorized travel. Most rural and hinterland inhabitants have to travel on average 1 and 4 hours, respectively.
This disparity is severely increased when using walking-only transportation, where even the suburban population is on average more than 3 hours away from the closest market. In the case of the average rural or hinterland inhabitant, the walking travel times needed to reach the closest market are 6 and 12 hours, respectively.
All accessibility metrics in Fig. \ref{fig:RurUrb_distr} reveal a clear pattern of class stratification: as rurality increases, from urban to hinterland areas, accessibility consistently declines while the travel times increase. Table \ref{tab:table_averages} reports in detail the average accessibility for the different urbanization levels for the five main African regions as defined by the UN \cite{UN_M49}. Results show that Middle Africa, the subregion with the highest share of population in rural and hinterland areas, ranks last in many of the urbanization classes in both motorized and walking travel times, as well as in entropy and availability measures when focusing on the most remote regions. At the same time, West Africa, the most populated subregion overall, has noticeably larger walking and motorized travel times in urban and suburban areas (more information in the  Supplementary Information section 3 Table 1 for subregional population per urbanization class).
This quantitative analysis emphasizes the critical scale of market accessibility challenges faced by hundreds of millions of individuals, especially in rural and remote areas.

\newpage
\subsection*{Market accessibility and economic status}
\label{RWI}
 
We employed the Relative Wealth Index (RWI) \cite{Chi2022} as a proxy for economic status (see \hyperref[subsec:RWI]{Methods} for a more detailed description of the RWI). 
Since the RWI is standardized within countries \cite{Chi2022, blumenstock2015predicting}, it does not allow for direct comparisons of economic disparities or absolute economic status across countries. For this reason, we examine the relationship between market access and the RWI within each country, focusing on second-level administrative unit resolution. 

\begin{figure}[ht!]
    \centering
    \begin{tabular}{l}
        \textbf{a}\\
        \includegraphics[trim = {8.4cm 0 9cm 0}, clip, width=0.85\linewidth]{ 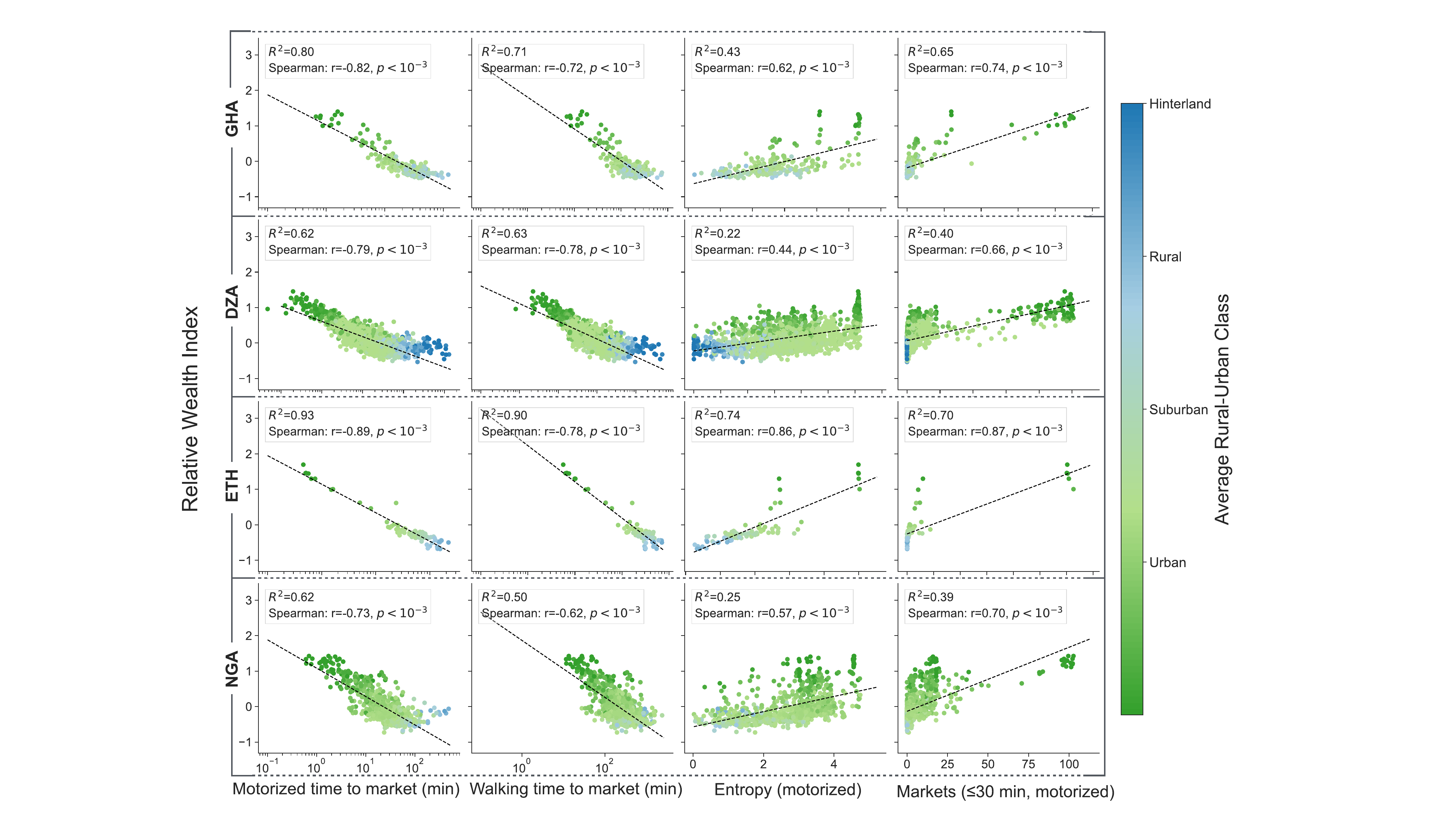}\hspace*{-2cm}\\
        \textbf{b}\\
        \includegraphics[trim={0cm 0 0cm 0}, clip,width=0.7\linewidth]{ 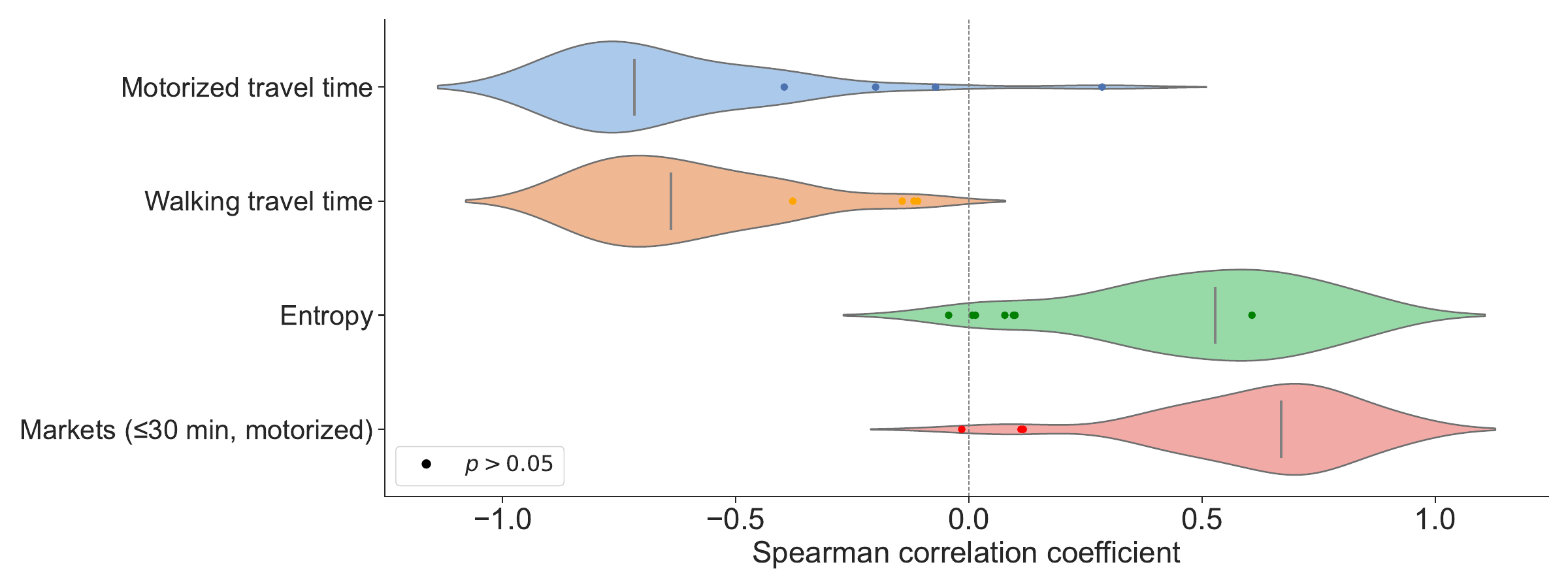}
    \end{tabular}

\caption{\textbf{Correlations between Relative Wealth Index and market accessibility.} 
\textbf{(a)} Correlations between the population averages of the RWI and the market accessibility metrics (motorized and walking travel times, entropy and number of markets within 30 minutes) at second-level administrative unit resolution for the countries of Ghana, Algeria, Ethiopia and Nigeria (coded in the figure as GHA, DZA, ETH and NGA). We also color the points with the population average of the rurality index to show the three-way relationship between market access, rurality, and relative wealth. For the travel times, we have set the x-axis as logarithmic.
\textbf{(b)} Distribution of Spearman correlation coefficients between the population averages of the RWI and the four accessibility metrics across 46 African countries. Correlations are studied at second-level administrative unit resolution within each country. Statistically significant correlations ($p < 0.05$) are shown for 42 countries. Non-significant values ($p > 0.05$), indicated as individual points, are typically associated with countries that have a limited number of second-level administrative units. Vertical lines represent the median correlation coefficient across all countries for each accessibility metric. Results in \textbf{(a)} and \textbf{(b)} highlight a significant rank correlation, indicating that wealthier regions present better market access and vice versa.}
    \label{fig:corr_RWI}
\end{figure}

The RWI has been rescaled to align with the spatial resolution of the WorldPop and market access rasters (see \hyperref[subsec:RWI]{Methods} for details), hence, analyzing at the second-level administrative scale helps reduce the noise that would otherwise result from pixel-level discrepancies.
Figure \ref{fig:corr_RWI} \textbf{b} presents the Spearman correlation coefficients for each of the 46 countries analyzed. 

Our findings reveal that wealthier regions (higher RWIs), which are typically more urbanized and economically connected, exhibit substantially shorter motorized and walking travel times, indicating better market accessibility, while economically disadvantaged areas consistently experience prolonged travel times, with limited numbers of markets reachable within a reasonable time.

Rural areas are economically less advantaged with respect to urban centers. Figure \ref{fig:corr_RWI} \textbf{a} presents four example cases from Ghana, Algeria, Ethiopia, and Nigeria portraying the three-way relationship between the RWI, market access, and rurality. Economic and spatial accessibility are strongly correlated, with households in regions of lower economic status, as measured by the RWI, often experiencing limited spatial access to markets. As an example, on average over all African countries, populations in the bottom RWI quintile face poor access to markets: their population-weighted mean travel time to the closest market is more than 1.5 hours by motorized transport, or 8 hours on foot. At the same time, they have, on average, only 2.4 markets within 30 minutes and an entropy of $1.5$, indicating limited spatial availability and poor distribution of nearby markets. These indicators capture different aspects of access: travel time is strongly affected by very remote pixels, while the count of markets within 30 minutes is bounded below at zero. In comparison, the average upper $20$th percentile within a country has to travel only around $10$ minutes by motorized transport or walk close to 45 minutes to reach the nearest market, with $130$ markets within 30 minutes and an entropy of $4.3$. 

Our results reveal a pattern consistent with studies in high-income countries, where disparities in food accessibility are strongly associated with socioeconomic and demographic factors \cite{block2004fast, kaufman1998rural, zenk2005neighborhood}. 

\clearpage

\subsection*{Market accessibility and food security}

We explored the relationship between market accessibility and food security, using the Integrated Food Security Phase Classification\cite{IPC_website} (IPC).
The IPC provides the proportion and total population in each of the five phases (1 to 5 in increasing food insecurity) over the analysis period. For our purposes, we aggregate IPC at first-level administrative unit resolution (see \hyperref[subsec: IPC]{Methods} for a more detailed description of the IPC). Since our spatial accessibility metrics are time-invariant and food insecurity presents high temporal variability, we used a five-year average of the IPC data to enhance robustness. 

\begin{figure}[ht!]
    \centering
    
    \includegraphics[width=0.9\linewidth]{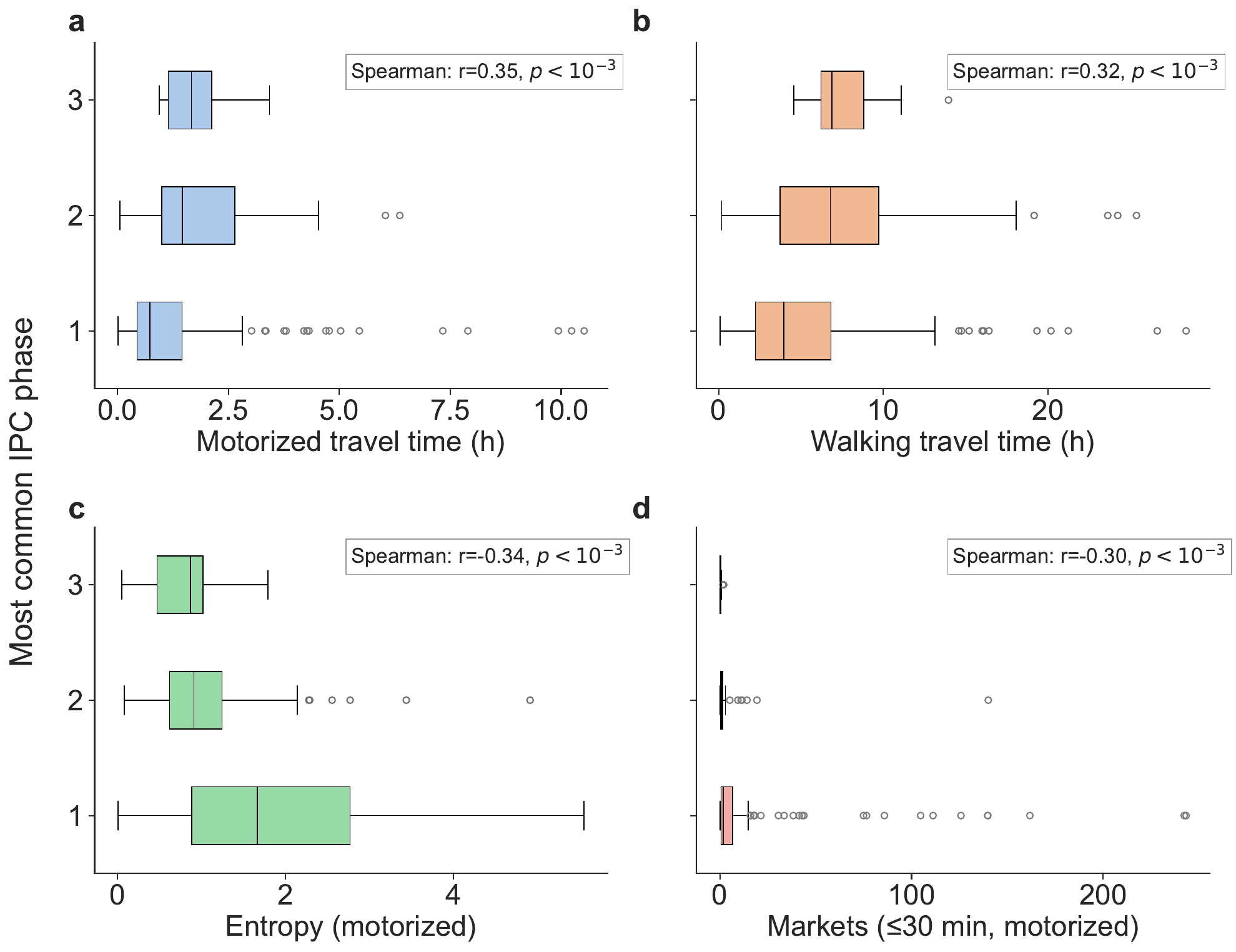}

    \caption{\textbf{Association between food insecurity and market accessibility.} Relationship between the most prevalent IPC phase and four market accessibility metrics at first-level administrative unit resolution across the African continent: \textbf{(a)} Motorized travel time, \textbf{(b)} Walking travel time, \textbf{(c)} Entropy of market distribution, and \textbf{(d)} Number of markets accessible within 30 minutes by motorized transport. Higher values in (a) and (b), and lower values in (c) and (d), indicate worse accessibility conditions. Each boxplot is based on the number of first-level administrative units classified under the corresponding IPC phase, with increasing values of IPC indicating more severe food insecurity levels. Box boundaries represent the interquartile range (IQR) between the first (Q1) and third (Q3) quartiles, while whiskers extend to the most extreme data point within 1.5 times the IQR from Q1 and Q3. Spearman correlation coefficients are reported for each accessibility metric, while all are statistically significant with $p<0.001$. }
    \label{fig:corr_IPC}
\end{figure}

Figure \ref{fig:corr_IPC} displays the distribution of population-weighted market accessibility values, averaged at first-level administrative unit resolution, across different IPC phases. Each region is assigned to its most prevalent IPC phase and shown against the region's population-average market accessibility.

Results show that less food-secure regions consistently exhibit worse market accessibility across all measured metrics. In IPC Phase 3 (Crisis), median motorized travel times, Fig. \ref{fig:corr_IPC} \textbf{a}, reach  1.6 hours, compared to 43 minutes in Phase 1 (Minimal/None). Walking travel times, panel \textbf{b}, follow a similar pattern, with median values of almost 7 hours in Phase 3 and 4 hours in Phase 1. Entropy values, Fig. \ref{fig:corr_IPC} \textbf{c},  capturing the evenness of market distribution, are notably lower in food-insecure areas, indicating more fragmented and uneven access. The median entropy in Phase 3 is 0.86, compared to 1.66 in Phase 1. Similarly, the number of markets accessible within 30 minutes by motorized transport, panel \textbf{d}, declines sharply, from a median of 1.80 in Phase 1 regions to 0.39 in Phase 3.

Overall, all accessibility metrics show significant rank correlations with IPC phase at the administrative level. These associations are confirmed by statistically significant yet moderate Spearman correlation coefficients ($p < 0.001$): 0.35 for motorized travel time, 0.32 for walking time, –0.34 for entropy, and –0.30 for cumulative market access. While these results underscore the association between food security and market accessibility, they also highlight the multifaceted nature of food insecurity, driven by a combination of factors such as economic conditions, price volatility, climatic variability, and conflicts. These findings are consistent with key studies in the literature, which, however, often focus on smaller-scale geographic analyses \cite{Bonuedi2022, Fraval2019}. Thus, while market accessibility is statistically associated with food security, it should not be interpreted as a standalone determinant, and its causal role requires deeper exploration in future research.

\section*{Discussion}\label{sec12}

In alignment with the ``leave no one behind'' principle, a core tenet of the UN's 2030 Agenda for SDGs, we identify spatial food market access gaps in Africa, contributing directly to SDG 2 (zero hunger) and SDG 10 (reduced inequalities). 
Our analysis introduces and compares three complementary accessibility metrics, namely, travel time to the nearest market, cumulative market availability, and entropy-based spatial distribution of markets. These three metrics demonstrate strong internal consistency, capturing distinct yet complementary aspects of spatial market accessibility. 

Our continental maps reveal a spatial market environment characterized by high accessibility in large urban centers and substantially lower accessibility in rural and hinterland areas: 360 million urban residents can reach more than 50 markets in less than 30 minutes by motorized travel, whereas 250 million rural and hinterland residents often face one- to four-hour journeys for a single option.

These spatial disparities can be interpreted through Turner et al.’s \cite{turner2018concepts} food-environment framework, which distinguishes between \textit{external} and \textit{personal} domains of access. Within this framework, our travel-time and cumulative-opportunity indicators capture structural dimensions of physical accessibility and availability in the \textit{personal} and \textit{external} domains respectively, while entropy refines this perspective by quantifying the evenness of accessible market distribution. In this sense, entropy reflects the degree of spatial redundancy in market options. High-entropy clusters, for instance in Cairo, Kinshasa, and Nairobi, are therefore more than simply dense concentrations of markets; they represent food environments where multiple, spatially dispersed options coexist, potentially enhancing robustness of access in the face of localized disruptions such as market closures or road failures.

All metrics confirm that urban areas consistently show higher accessibility, while rural and hinterland areas, home to a substantial portion of Africa’s population, exhibit average travel times that exceed several hours to a limited number of nearby markets. 
These spatial patterns are closely linked to socioeconomic conditions. Using the Relative Wealth Index (RWI) as a proxy for socioeconomic status, our analysis shows that the poorest national quintile travels, on average, more than 1.5 h by vehicle (7 h walking) to the nearest market, against 15 min (1 h walking) for the richest quintile.
Our findings echo literature from high-income contexts, emphasizing how spatial and economic marginalization often overlap to exacerbate food access challenges \cite{lang1998access,Larsen2008,Short2007}. 
This is a critical twofold issue since economically disadvantaged regions not only face financial barriers to accessing food markets but also suffer from limited spatial accessibility, compounding their food security challenges. 

Spatial remoteness is not merely an inconvenience; it inflates both direct and opportunity costs (time, fuel, labor) \cite{olvera2013puzzle} and has been linked empirically to lower food security \cite{bonuedi2022agricultural}.
In fact, the moderate yet significant correlation between accessibility and IPC ($\rho \approx 0.32 - 0.35$) suggests that areas with higher IPC (i.e., more food insecurity) consistently show longer travel times and reduced market access. We stress that this relationship is purely associative and should not be read as evidence of causation, rather as a reflection of the multifactorial nature of food security, which is also influenced by external factors such as climate shocks, conflict, and market prices
\cite{martin2019food, niles2018cross, headey2016impact}.

We acknowledge several limitations that temper our conclusions. Regarding the data, OSM coverage can be sparse and uneven across African countries, particularly in rural areas where volunteered geographic information is collected less frequently. Consequently, some portion of the observed rural–urban accessibility gradient may reflect differences in market mapping completeness and intensity in addition to genuine disparities in physical accessibility. Although WFP data substantially mitigates these gaps—especially in conflict-affected, food-insecure, and remote regions (see \hyperref[subsec:WFP]{Methods} )—its coverage remains incomplete, with 12 out of 54 African countries not represented at the time of publication (e.g., Angola, Botswana, and Namibia). Such heterogeneity may lead to an underestimation of market density in sparsely mapped regions. Importantly, however, the travel time to the nearest market depends primarily on proximity to the closest mapped market rather than overall market presence, making it less sensitive to variations in mapping intensity than purely count-based measures. 

Beyond inconsistencies in static market data, rural food access frequently depends on informal farmer shops and ambulant markets, which compensate for limited permanent retail infrastructure and play a central role in smallholder exchange \cite{satyam2022resilience, Usman2021, Bonuedi2022}. Because these vendors operate intermittently and are spatially mobile, they are not consistently represented in large-scale geospatial datasets. As a result, our analysis captures spatial access to formalized and mappable market nodes, rather than the full diversity of food acquisition channels present in rural contexts. Overall, the findings should be interpreted as capturing relative structural patterns of accessibility under incomplete information, highlighting the need for more consistent, large-scale, and systematically collected datasets of food markets across the continent.

In parallel to the limitations on the market data, the accessibility metrics rely on travel time estimates, which can be unreliable due to several factors, such as missing walkable paths, commonly used routes, or detailed road-specific information, that influence travel speeds within the precomputed friction surface utilized (see \hyperref[subsec: fric_surf]{Methods} for more details on travel time computation). Additionally, the friction surfaces used to compute travel times date back to 2020, whereas the OSM and WFP data were last extracted in December 2024. As a result, recently constructed or newly mapped roads are not included in the travel time computations. For this reason, we performed a validation analysis comparing travel times to markets using the friction surface and the latest road network available in OSM (see Supplementary Information section 8), where we found strong positive associations between the two approaches, supporting the robustness of our travel-time estimates. Finally, there is a lack of detailed and reliable market-specific information in the OSM data, such as size, food prices, and food variety, which is essential and could be included in the future for a more nuanced analysis of market functionality under real-world conditions.

Despite these limitations, our analysis provides a scalable foundation for future integration of higher-resolution market datasets and offers stakeholders and humanitarian actors a data-driven framework to identify food market accessibility gaps and inform context-sensitive assessments related to inequalities. The spatial patterns revealed in this work intersect with an ongoing policy debate on the role of public versus private actors in food retail provision. While many policymakers increasingly favour private retail development to reduce fiscal, administrative, and liability burdens \cite{reardon2003rise,resnick2020politics}, traditional and publicly governed markets remain critical food distribution nodes, particularly for low-income and rural populations, where private retail expansion is often limited \cite{davies2022governance,cook2024nutritional}. In this context, our accessibility metrics could provide a spatial lens to assess the inclusiveness and spatial configuration of market access, especially in resource-constrained and food-insecure settings.

From an operational perspective, the food market accessibility metrics presented in this paper can provide a robust evidence base to support decision-making in multiple ways. First, they provide a scalable tool to identify subnational areas where populations face structurally high travel times and limited market availability, enabling prioritization of infrastructure investments such as road upgrades or improved market placement. Second, they can support humanitarian planning by highlighting locations where poor physical access may exacerbate vulnerability during food system disruptions, including price shocks, climatic events, or conflict-related supply interruptions. Finally, while the proposed indicators are time-invariant, they can serve as baseline structural predictors within early-warning or forecasting frameworks \cite{voukelatou2026predicting, herteux2024forecasting} when combined with time-varying inputs (e.g., rainfall anomalies, vegetation indices, market prices, or displacement data), helping to identify regions where shocks are more likely to translate into severe access constraints.

\section*{Methods}
\addcontentsline{toc}{section}{Methods}
\phantomsection
\label{sec:methods}

This section describes the data used in our analyses and the methodologies applied to harmonize market definitions across the OSM and WFP datasets. We also detail the three accessibility metrics used in this study: cumulative accessibility, travel time to the closest market (both motorized and walking), and entropy (see the Supplementary Information section 4 Table 2 for a summarized view of the strengths and limitations of the metrics).\\

\textbf{OpenStreetMap data}\\

OSM is a collaborative open-source project providing free geographic data contributed by volunteers. It includes detailed road networks (type, surface, direction), buildings, commercial establishments (shops, markets, services), land use, and points of interest. Given the focus of the study on food markets, we extracted the OSM shops and amenities tagged as food-related \cite{OSM_Wiki_Food_2024} for all African countries. Specifically we considered the following amenities: \textit{Marketplaces, Supermarkets, Bakeries, Butchers, Convenience Stores, Dairy Shops} and \textit{Farm Shops}. 
For each one of these, we extract the geographical coordinates. 
Since OSM relies predominantly on volunteer contributions, it tends to have less coverage in LMICs or conflict-affected areas, which often remain underrepresented.\\

\textbf{WFP market data}
\addcontentsline{toc}{section}{WFP}
\phantomsection
\label{subsec:WFP}\\

WFP's Market price and Market Functionality Index data \cite{wfp_market_1990_2025, wfp2020mfi} covers 42 out of the 54 African countries, and while having fewer market entries than the OSM dataset (See Supplementary Information Fig. 3), it provides food market locations in LMICs, covering conflict-affected, food insecure and remote regions. As a result, WFP's dataset helps mitigate gaps in OSM coverage. In the Supplementary Information Fig. 4, we illustrate an example of two African countries, Ethiopia (left) and Nigeria (right), to highlight how WFP data complements OSM, especially in rural and high-vulnerability contexts.
In the Supplementary Information section 5 Fig. 3, we present the contribution of both OSM and WFP data at a country level.

We compute the spatial overlap between both datasets at 200 m or less for all of Africa. A total of $404$ out of $3380$ WFP markets ($11.95\%$) overlapped with at least one OSM point, and the average overlap ratio over all countries is $14.10\%$. The low overlap observed between OSM and WFP data demonstrates how the two datasets complement each other, increasing the quantity and geographic coverage of our study. 

We emphasize the fundamental differences in nature and definition between the OSM and WFP datasets. In OSM, each data point represents an individual amenity. In contrast, a WFP data point represents a market with multiple amenities, which might not be located at the same point but function collectively as a unified market. In other words, a WFP market represents a physical space where buyers and sellers engage in trade \cite{WFP_MFI}. To account for the differences in the definition of market between the two data sources, we follow the methodological steps presented in the \textit{Food market definition} section.\\

\textbf{Demographic data}\\
\addcontentsline{toc}{section}{WPP}
\phantomsection
\label{subsec:WPP}

To account for population density, we utilize data from WorldPop \cite{WorldPop_UNadj_Unconstrained_2000_2020}, which offers high-resolution, open-access population distribution estimates. WorldPop integrates census data, household surveys, satellite imagery, and other geospatial sources to produce detailed demographic datasets. Specifically, we use population count data aggregated in a raster format at a 30 arc-second resolution (approximately 1 km at the equator) from the latest available year, 2020. In this dataset, the population counts are adjusted to match the official United Nations estimates and are unconstrained, meaning that population is statistically redistributed across all grid cells without restricting settlement to plausibly inhabited land.  Each raster cell contains an estimate of the absolute number of inhabitants residing within it.\\

\textbf{Socioeconomic data}\\
\addcontentsline{toc}{section}{RWI}
\phantomsection
\label{subsec:RWI}
To incorporate the socioeconomic narrative to market accessibility, we use the Relative Wealth Index (RWI) \cite{rustein2004dhs}, an indicator of household relative wealth developed by the Demographic and Health Surveys (DHS) Program. This index can take positive or negative values depending on whether the household is wealthier or less wealthy than the rest in the region. This index is computed using the principal component analysis procedure and is available at the household and regional levels. 

In our case, since we need a finer resolution, we use the RWI estimates in LMICs generated from Meta \cite{Chi2022}. These estimates consist of predictions of RWI in a rasterized form, using machine learning models trained on satellite imagery and survey data. The lower-scale resolution enables a more detailed spatial analysis of economic disparities. The raw data is provided in the Bing Maps Tile System \cite{BingMaps_TileSystem} format. We rescale the raster with the standard averaging method from Python's library \textit{rasterio} to be consistent with the population raster resolution of 30-arc seconds.

We conduct a rural-urban comparison of market accessibility using data from the Global Urban Rural Catchment Areas (URCA) dataset \cite{Cattaneo2020}. This dataset includes a raster map of the 30-category urban-rural continuum, indicating catchment areas around cities and towns by travel time \cite{Cattaneo2020}. We used a simplified 4-level categorisation of the dataset, namely the urban, suburban, rural, and hinterland classes in increasing order of remoteness. The pixels labelled in the urban class are in a city of more than 20 thousand inhabitants, while suburban pixels are less than 1 hour away from such a city. Rural and hinterland areas are considered to be within 1-2h and $\geq$ 2h from a city of 20 thousand inhabitants, respectively. For a detailed comparison between this classification and the official United Nations Population Division estimates \cite{un_wup_2025}, see the Supplementary Information section 3. \\

\textbf{Food security data}
\addcontentsline{toc}{section}{IPC}
\phantomsection
\label{subsec: IPC}\\

To assess the correlation between market access and food security, we use the Integrated Food Security Phase Classification (IPC) for Acute Food Insecurity \cite{IPC_website}, a standardized tool used to assess and classify the severity of acute food insecurity in a region. It provides a consistent and comparable framework to analyze food security conditions, identifying populations in need and the underlying drivers of food insecurity. The IPC uses a five-phase scale: (1) Minimal/None, (2) Stressed, (3) Crisis, (4) Emergency, (5) Catastrophe/Famine. This index is widely used for decision-making in humanitarian and development interventions.

The IPC data provides both the proportion and total population in each of the five phases during the analysis period. For the purposes of the study, we use the data aggregated at first-level administrative unit resolution. IPC data is temporal and may be generated multiple times per year, depending on the country. Since our accessibility metrics are time-invariant, considering they are constructed from the most recent OSM and WFP, we average the IPC dataset over the last five available years for each country. Notably, the most prevalent phase at first-level administrative unit resolution is highly skewed toward phase 1, resulting in an extreme class imbalance (see Supplementary Information section 7).\\

\textbf{Food market definition}\\
\addcontentsline{toc}{section}{kmeans}
\phantomsection
\label{subsec: kmeans}\\
Since OSM data consists of individual food shops that differ in the type of products they sell, we adopt a rough market definition provided by WFP, where a market is defined as a small commercial area containing several food shops \cite{WFP_MFI}. We adapt the OSM data to WFP's definition by clustering data points within a 15-minute walking distance. To achieve this, we employ an iterative \textit{K-means} clustering algorithm, increasing the number of clusters until meeting the condition that the maximum cluster radius does not exceed 15 minutes of travel at the average walking speed of $1.42 \: \text{m/s}$ (a total distance of approximately $1300 \text{m}$). The centroids of these clusters are then designated as market locations (see the Supplementary Information section 6 Fig. 5 \textbf{a} for an example portraying the data clusters in Addis Ababa and a sensitivity analysis on the cluster radius, Fig. 5 \textbf{b}). The harmonized market dataset is publicly available in the repository linked in the Code Availability section.
\\

\textbf{Friction surfaces and travel time to the closest market}

\addcontentsline{toc}{section}{Friction Surface}
\phantomsection
\label{subsec: fric_surf}

Travel time to the nearest facility is one of the most widely used spatial measures of accessibility in the literature, particularly in the domain of food accessibility, where analyses often rely on survey data. Despite its simplicity and inherent limitations, such as the assumption that individuals always travel via the shortest route and access the nearest facility, its widespread use can be attributed to its intuitive interpretability, making it especially effective for communication with relevant stakeholders and decision-makers. 

Advances in Geographic Information System (GIS) technologies and the growing availability of spatial data have enabled accessibility studies to expand across multiple scales, ranging from city-level analyses to national and global assessments. These studies have mapped the shortest travel times to key locations, such as healthcare centers or urban hubs \cite{Weiss2020, Palk2020, Weiss2018}, providing valuable insights into spatial accessibility patterns. 

Methodologically, the estimation of travel time typically relies on shortest-path routing through street networks and/or public transport systems \cite{higgs2004literature}. To reduce computational complexity in larger areas, isochrone calculations are also a common approach in the accessibility field \cite{o2000using, xi2018exploring}. These approaches, however, depend heavily on the accuracy and completeness of road network data. Recognizing that transport in many rural areas of Africa may occur outside mapped roads or trails, we adopt a more flexible approach commonly used in the GIS community, friction surfaces. A friction surface is a raster dataset that encodes travel speed for each pixel based on a combination of factors, including roads, transport networks (e.g., railways and navigable waterways), land cover types, bodies of water, topographic conditions (elevation and slope), and national borders \cite{Weiss2020, Weiss2018}. This method allows for travel time estimation even in areas where roads are absent, as all pixels, regardless of mapped infrastructure, are considered for routing based on terrain conditions. This flexibility makes friction surfaces particularly well-suited for modeling travel times in data-scarce and rural contexts.

In this work, we use two friction surface rasters (walking and motorized) from \cite{Weiss2018} at the 30 arc-seconds scale. Using this method, the travel time from each origin (pixel) to its closest market can be easily computed using the standard algorithms for shortest path-finding. In our case, we used the \textit{MCP\_Geometric} algorithm from \textit{scikit-image}'s \textit{graph} module \cite{van2014scikit} in \textit{Python}.
Our scope is limited to journeys within each country; cross-border travel falls outside the present analysis.

The three accessibility metrics employed in this study require an estimation of the travel time to the food markets. Having disclosed travel time to the closest market, in the following sections, we provide a more detailed description of the two remaining accessibility metrics.\\

\textbf{Availability}\\

To quantify the availability of markets, we used the framework of the cumulative accessibility \cite{hansen1959accessibility}, in which a catch radius given by a time budget ($T$) is established around the centre of a given origin region $i$. The total number of markets inside this area gives the accessibility of region $i$. Mathematically, it can be expressed using the Heaviside step function ($H(t_{ij})$), where $t_{ij}$ is the travel time from the origin pixel $i$ to each market $j$.

\begin{equation}\label{eq:cumul_acc}
    A_i = \sum_j (1-H(t_{ij})) \quad ; \quad H(t_{ij}) := \begin{cases} 1, & t_{ij} \geq T \\
                     0, & t_{ij} < T
       \end{cases}
\end{equation}

We note that other more complex forms of cumulative accessibility exist, such as the Hansen type measure, \cite{hansen1959accessibility},  which considers a decaying function (usually in gaussian or negative exponential forms) to modulate the relevance of travel cost or the gravity-based Shen-type measure \cite{shen1998location}, which accounts for supply and demand ratios and has evolved into the various 2-Step Floting Catchment Area (2SFCA) methods \cite{luo2003measures}. 

Despite the improved completeness of these metrics, they bring certain limitations, such as the need for supply data (e.g., market capacity or size) and the reliance on parameter tuning, which often depends on mobility data that is largely unavailable in developing countries. Moreover, the increased complexity of these metrics reduces their explainability. Therefore, we adopt the simplest form of cumulative accessibility, as we believe that in data-scarce contexts, simpler measures yield more reliable and consistent results while adding complexity leads to approximations, estimations, and heuristics that could be far from reality.

In our work, we used a 30-minute threshold in motorized travel. One could rightly argue that changing this time could notably alter the results. Indeed, this value can be tuned to better suit the scale of the analysis or the potential area of influence of the market users. To assess the consistency of the results, we performed a sensitivity analysis of the cumulative accessibility in Ethiopia by looking at the evolution of the population average of the cumulative accessibility when exploring different values of the time threshold (see Supplementary Information section 2 \textbf{b} and \textbf{c} for sensitivity plots). Overall, the cumulative accessibility remains stable when changing the threshold in the vicinity of 30 minutes, supporting the robustness of the proposed value.\\

\textbf{Utility accessibility: entropy}\\

Utility accessibility \cite{Handy1997} is based on the probability of a user choosing a facility. This probability depends on the ``utility'' an individual assigns to each facility, which may depend on the socioeconomic status of the individual as well as his subjective perception. Utility accessibility is a more complex measure than the previously described, which reduces its interpretability.

Since we focus on spatial market accessibility, we propose a probability of choosing a market based solely on travel time.

Consider a populated area $i$. Consider also a set of $m$ target marketplaces ${j_1, j_2, ..., j_m}$ at travel times ${t_{ij_1}, t_{ij_2}, ..., t_{ij_m}}$ from the populated area. Given this framework, we assume that the inhabitants of $i$ will access the markets according to a probability that decays with travel time to each one of them. In other words, closer markets will have a larger probability of being visited than distant ones. For consistency with the literature, we have set this probability distribution as a normalized version of the exponential impedance function commonly used in the Shen- and Hansen-type accessibility metrics \cite{shen1998location, hansen1959accessibility}, which resembles the Boltzmann distribution used in the framework of statistical physics.

\begin{equation}\label{eq:boltz_prob}
    p(i\rightarrow j|i) = \frac{e^{-\beta t_{ij}}}{\sum_k e^{-\beta t_{ik}}},
\end{equation}

In Eq. \ref{eq:boltz_prob}, $\beta$ is the sensitivity parameter (inverse temperature) of the inhabitants. According to this probability distribution, if the sensitivity is high, the most visited marketplaces will be the ones closest to the populated centre, whereas, for low sensitivities, near or distant markets will have similar visiting probabilities.

Once the probability density is defined, we can compute the different quantities related to the mobility behaviour of the users and the spatial distribution of the markets. Firstly, we can obtain the average travel time of the individuals in $i$ as:

\begin{equation}\label{avg_dist}
    \left<t_i \right> = \frac{\sum_j t_{ij}e^{-\beta t_{ij}}}{\sum_j e^{-\beta t_{ij}}},
\end{equation}

Secondly, we may study the entropy of the distribution and obtain a proxy for the accessibility

\begin{equation}\label{entropy}
    S_i = -\sum_j p(i\rightarrow j|i) \log(p(i\rightarrow j|i)).
\end{equation}

As described in a similar approach defined in \cite{Erlander1977}, the entropy measures the spread of the distribution of journeys among all the target facilities, and thus, the extent of markets used by the population in a given area. Intuitively, if the spatial distribution of markets in an area is even (usually many close-by facilities) the entropy and thus the accessibility will be large, whereas if the distribution is biased towards a reduced set of closer facilities (therefore lower accessibility), the entropy will be low. In this case, since the probabilities depend solely on travel time, the entropy informs about how homogeneous the perceived travel time distribution is around a given area (perceived with the probability defined in Eq. \ref{eq:boltz_prob}). Therefore, in our case, entropy can be considered a measure of accessibility describing the spatial distribution of facilities.

One of the issues with using raw entropy as an accessibility metric is that a homogeneous distribution of large travel times can also produce larger values of entropy. Acknowledging this limitation, we define an upper bound for travel time above which the market would be deemed inaccessible ($p(i\rightarrow j|i) = 0$). In our case, we set a large threshold of $t = 4\text{h}$ of motorized travel time to make sure we don't leave accessible markets out. Finally, we set the value of $\beta$ following the common approach for exponential factor used in the 2SFCA \cite{Pan2015}. Namely, we set the exponential factor to $0.01$ for a travel time of $60$ minutes, $e^{-\beta 60 (min)} = 0.01$. Then, we isolate the value of the sensitivity: 

\begin{equation}
    \beta = \frac{-log(0.01)}{60} \approx 0.077\text{min}^{-1}
\end{equation}

\subsection*{External comparison of travel-time estimates}
To assess robustness of travel-time estimates, we conducted an external comparison using the OpenRouteService API, which computes shortest-path travel times on the OpenStreetMap road network. We sampled 100 random origin–destination pairs (raster pixels to market locations), restricting the sample to cases where both endpoints were within 2 km of a mapped road to ensure network routability. We compared friction-surface-based travel times with network-based routing estimates for both motorized and walking modes. Results show a consistent positive association between the two approaches (see Supplementary Figures 7–8), with discrepancies attributable to differences in modeling assumptions and potential incompleteness of mapped road infrastructure.

\section*{Code Availability}

The code computing the accessibility metrics and generating the plots in this work is available in \url{https://github.com/robedamo/food_market_access.git}.

\section*{Data Availability}

Market locations were obtained from OSM's geofabrik file from Africa (\url{https://download.geofabrik.de/africa.html}), accessed on Dec. 2024. WFP markets data were retrieved from WFP's Market price and Market Functionality Index database (\url{https://dataviz.vam.wfp.org/economic/overview?current_page=1}), accessed on Dec. 2024. Friction surfaces were downloaded from the Malaria Atlas project (\url{https://malariaatlas.org/project-resources/accessibility-to-healthcare/}), while population counts come from WorldPop \cite{WorldPop_UNadj_Unconstrained_2000_2020}(\url{https://hub.worldpop.org/geodata/listing?id=75}). RWI data was obtained from the Humanitarian Data Exchange (HDX) portal (\url{https://data.humdata.org/dataset/relative-wealth-index}), accessed on Jan. 2025. Finally, we also retrieved the IPC data from the HDX portal, accessed on Jan. 2025, (\url{https://data.humdata.org/dataset/ipc-country-data}) and the rural-urban catchment areas from the URCA project \cite{Cattaneo2020} (\url{https://data.apps.fao.org/catalog/iso/9dc31512-a438-4b59-acfd-72830fbd6943}) 

\bibliography{references}

\section*{Acknowledgements}

We would like to thank Duccio Piovani, Frances Knight, Kyriacos Koupparis, Angela Di Perna, Ibrahima Mamadou Diarra, and Jonas De Meyer for their support and feedback.
RBD, DP, SF, RS, and KK acknowledge support from the Lagrange Project of the Institute for Scientific Interchange Foundation (ISI Foundation) funded by Fondazione Cassa di Risparmio di Torino (Fondazione CRT).

\section*{Author contributions statement}

\noindent \textbf{Conceptualisation}: RBD,  KK, VV; 
\textbf{Methodology}: RBD, RS, KK, VV, SF; 
\textbf{Formal analysis}: RBD;
\textbf{Investigation}: RBD, VV, KK; 
\textbf{Writing -- Original Draft}: RBD,  KK, VV; 
\textbf{Writing -- Review \& Editing}: all authors;
\textbf{Supervision}: KK.

\section*{Funding}
This work was supported by Fondazione Cassa di Risparmio di Torino (Fondazione CRT) and by the Bill \& Melinda Gates Foundation [grant number INV-037325].

\section*{Additional information}
\textbf{Competing interests} Authors declare no competing interests.

\end{document}


\title[Article Title]{Continental-scale assessment of spatial food market accessibility in Africa using open geospatial data}
\maketitle

\section{Cumulative accessibility and entropy using walking travel times}

\begin{figure}[ht!]
    \centering
    \begin{tabular}{ll}
         \textbf{a} & \textbf{b}\\
         
        \includegraphics[width=0.48\linewidth]{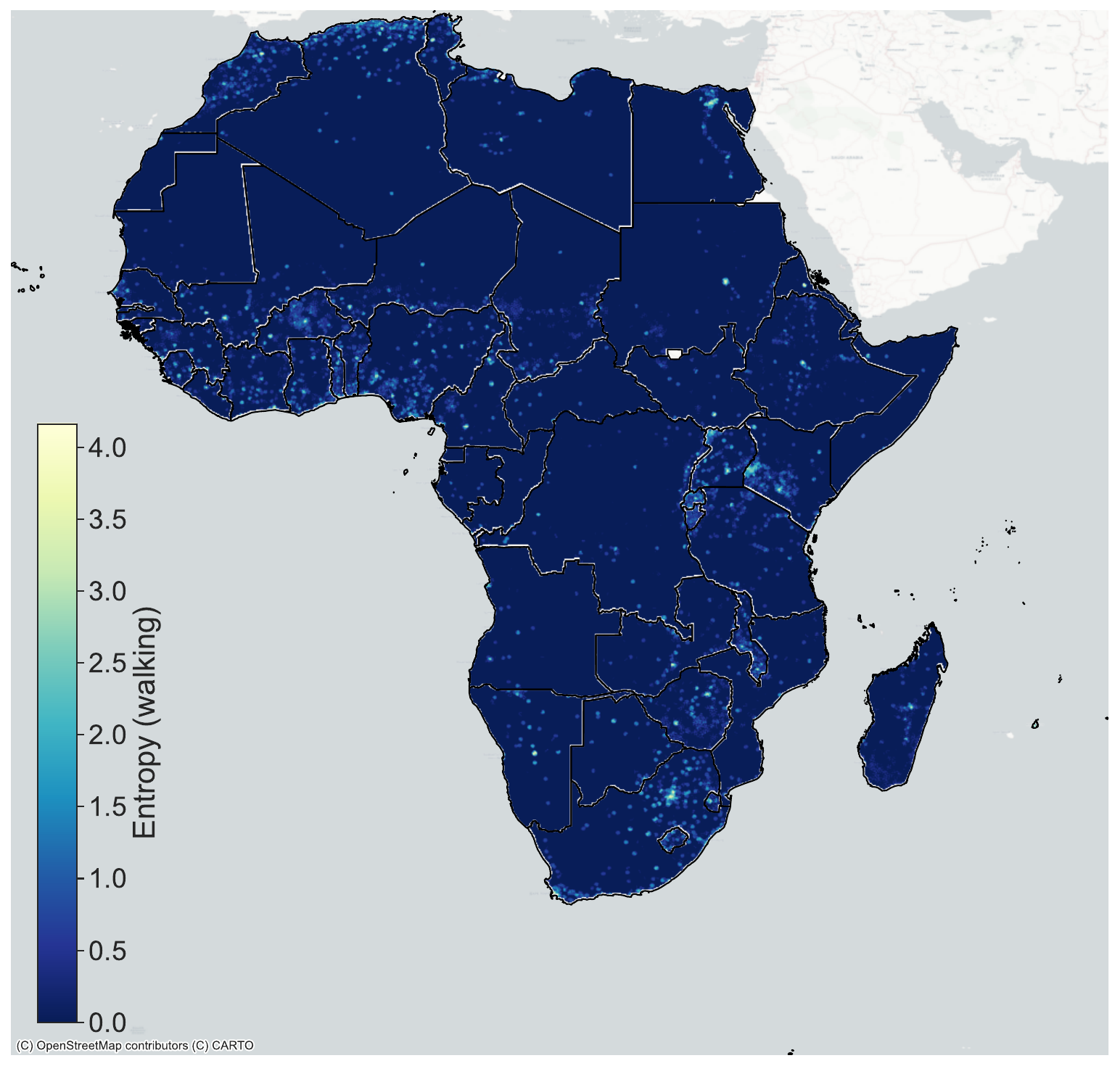}& 
        \includegraphics[width=0.48\linewidth]{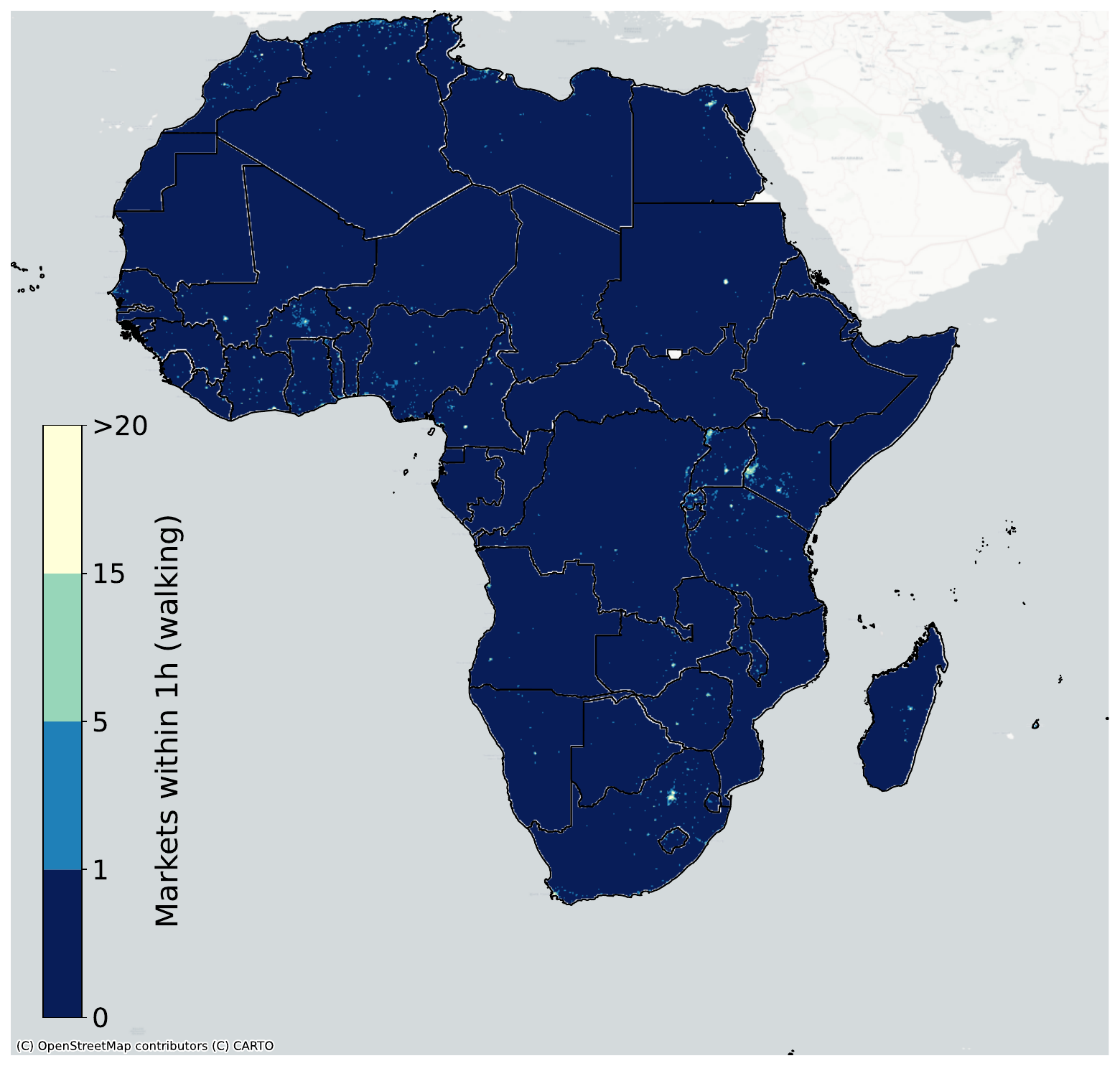}
    \end{tabular}

    \caption{\textbf{Walking‑only accessibility surfaces for entropy and number of markets}. Maps for Africa showing \textbf{(a)} Entropy market distribution and \textbf{(b)} Number of markets within 1 hour, both computed using the walking-only friction surface. Both panels use the same 30‑arc‑second ($\approx$1km) resolution as in the text.}
    \label{fig:walking_en_cumul}
\end{figure}

Because walking is the default mode of mobility for many rural African households, we recomputed cumulative accessibility (number of markets reachable within 60min) and entropy (spatial distribution of reachable markets) using a walking‑only friction surface. Results mirror those obtained for motorised travel, though accessibility values are markedly lower and inequality gradients steeper (see Fig. \ref{fig:walking_en_cumul}).

\section{Correlation between accessibility metrics and sensitivity analysis}\label{app:corr_mat}

\begin{figure}[ht!]
    \centering
    \begin{tabular}{ll}
        {\bf a} & {\bf b}\\

        \includegraphics[trim = {2cm 0 3.6cm 0}, clip, width=0.485\linewidth]{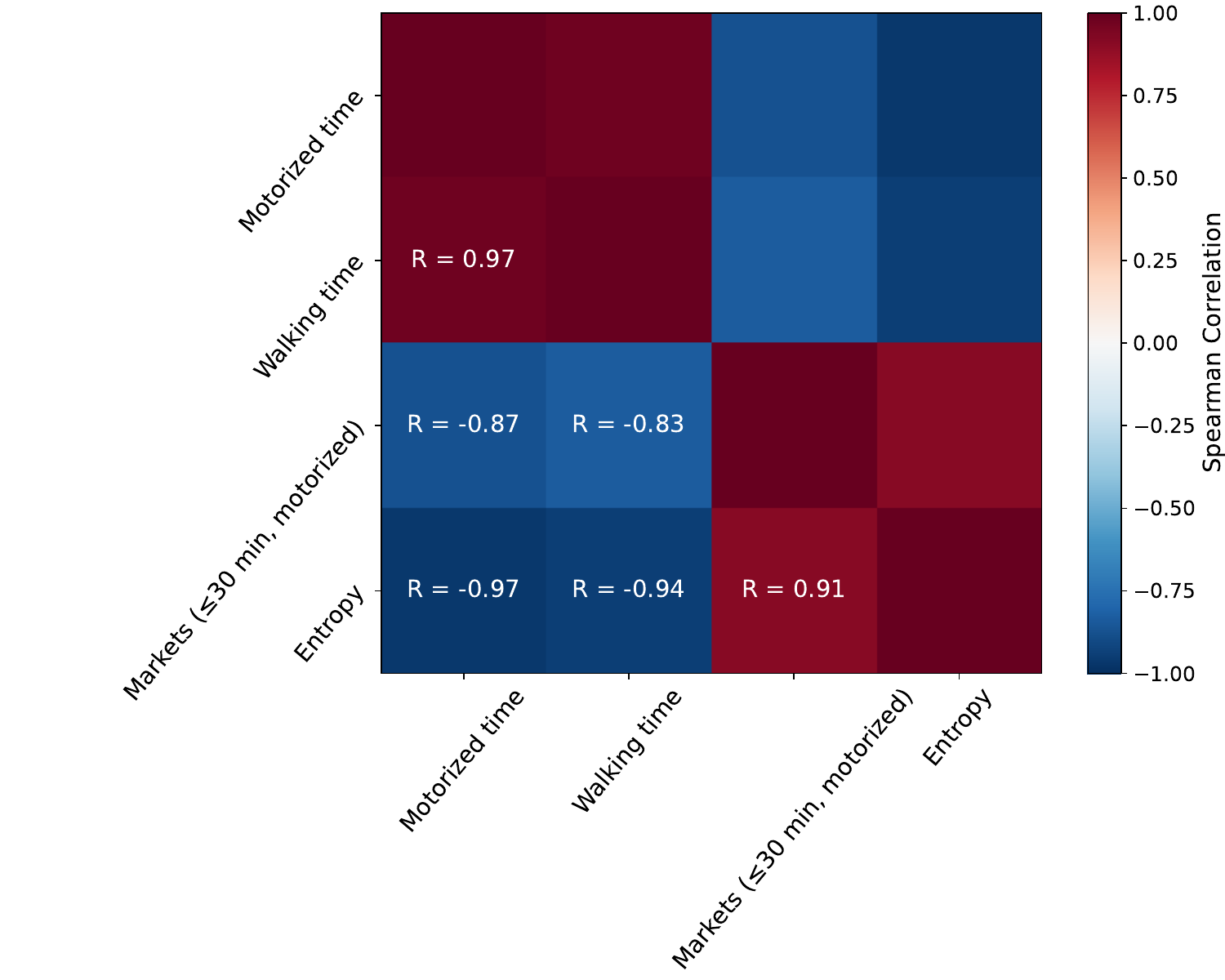} 
        &
        \includegraphics[trim = {7.3cm 0 0cm 0}, clip, width=0.445\linewidth]{ 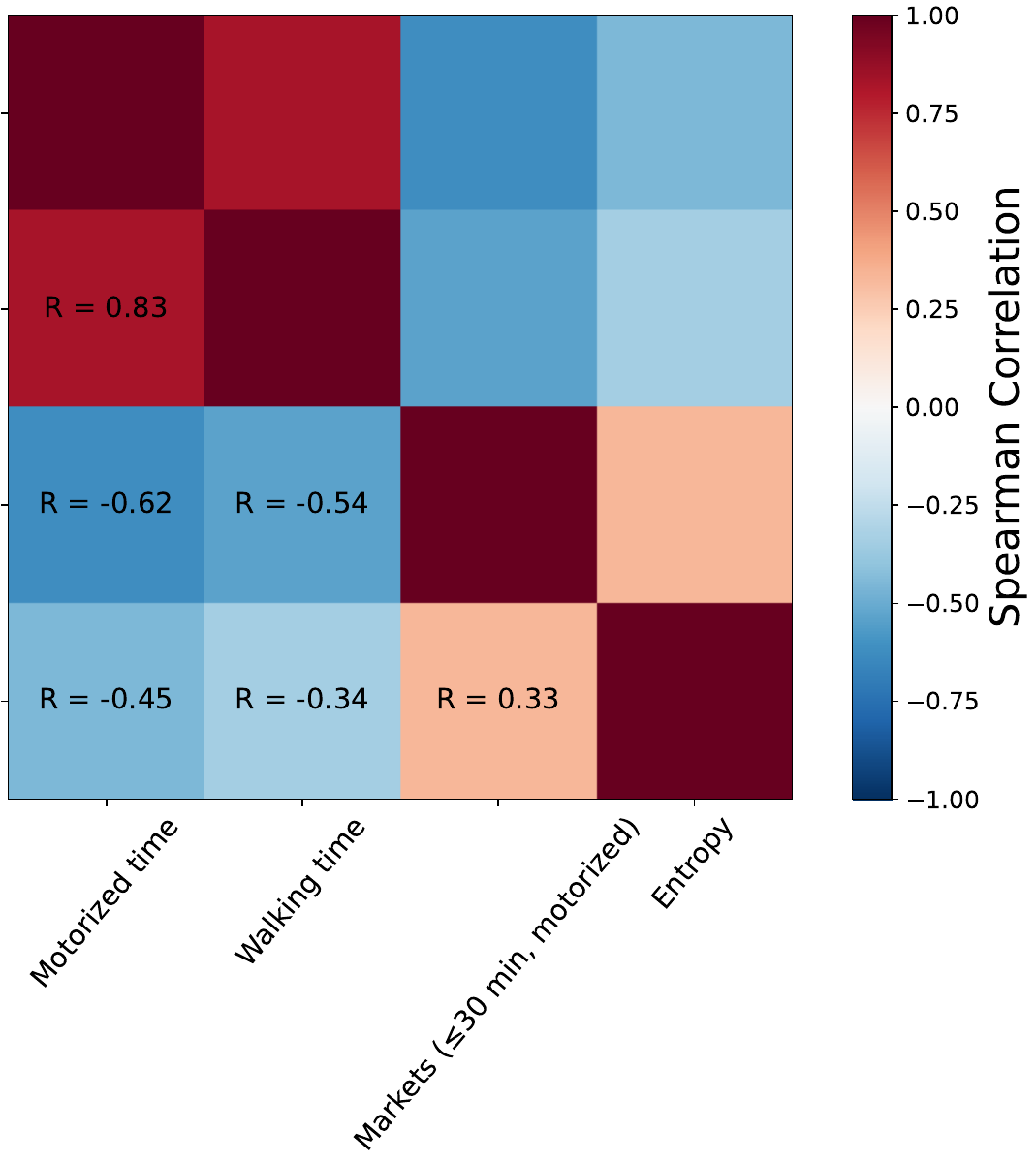}\\
        
        {\bf c} & {\bf d}\\

        \includegraphics[width=0.49\linewidth]{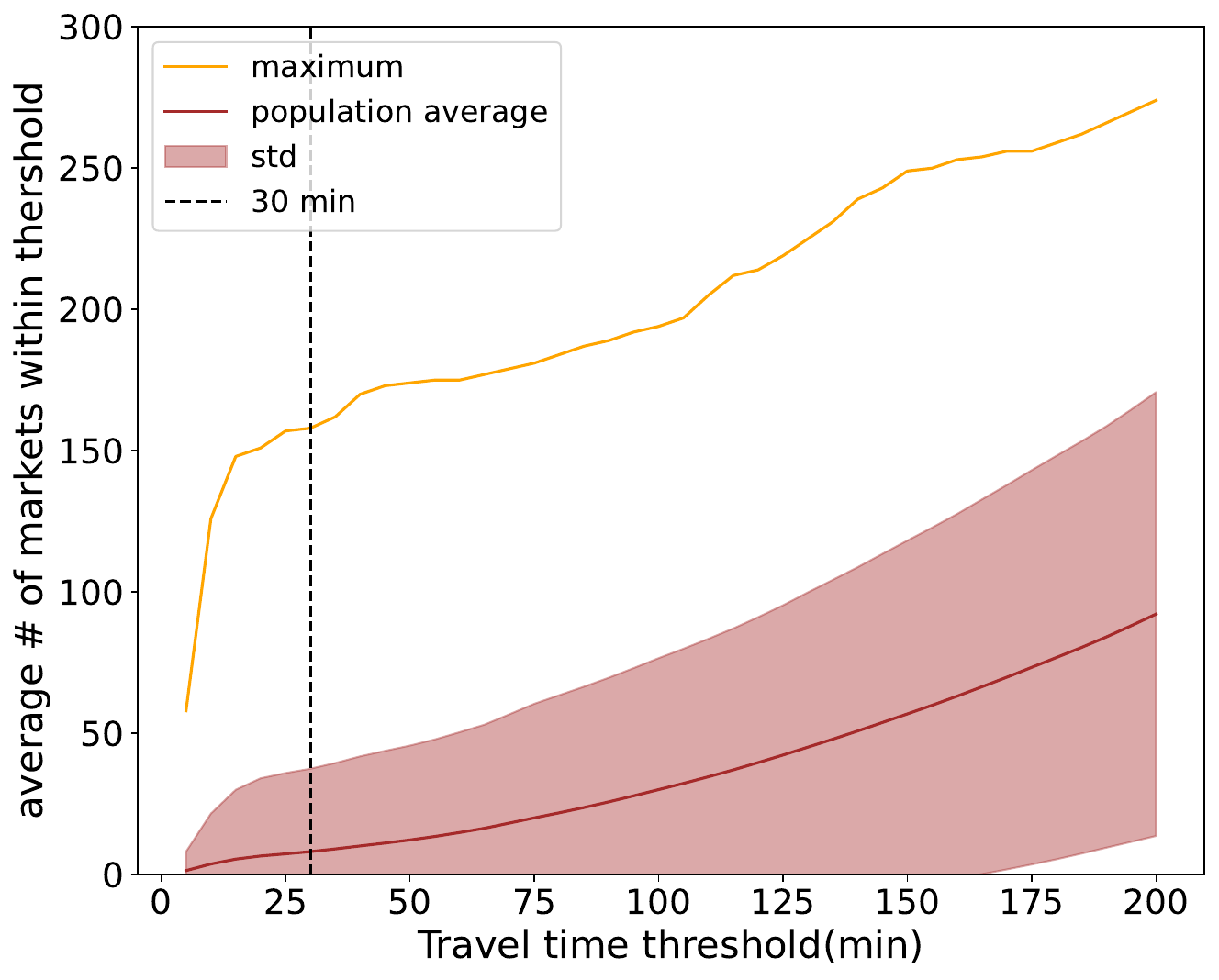}
        &
        \includegraphics[width=0.49\linewidth]{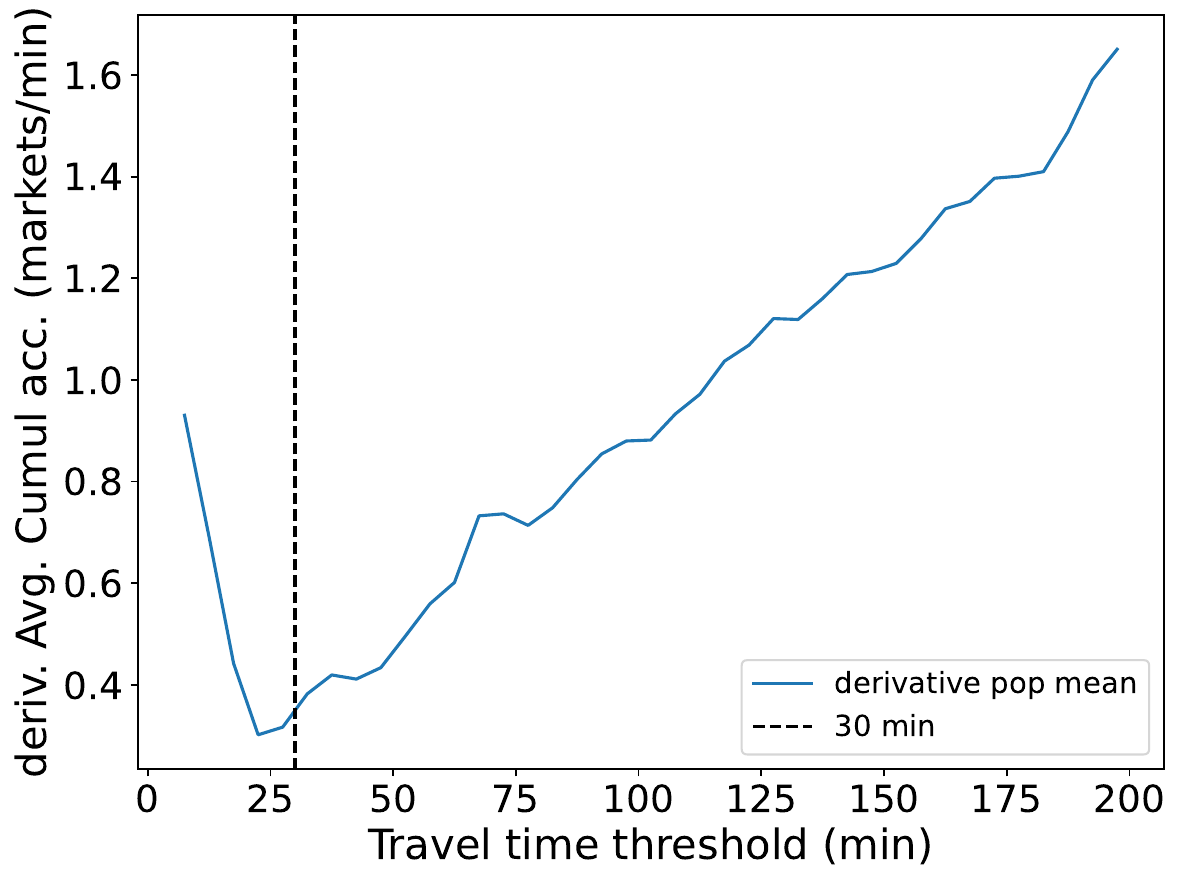}

    \end{tabular}
    
    \caption{\textbf{(a)} Matrix displaying the Spearman correlation between population-weighted median accessibility values at the country level. \textbf{(b)} Matrix displaying the Spearman correlation at the pixel level. For each country, we computed the rank correlation coefficient. The value displayed is the average of such coefficients across all African countries. \textbf{(c)} Evolution of different statistics of the cumulative accessibility with the threshold travel time value in Ethiopia. \textbf{(d)} "Derivative" of the average and population average of the cumulative accessibility with respect to the travel time threshold in Ethiopia. 
    }
    \label{fig:metric_correlations_sensitivity}
\end{figure}

To test the coherence of our three accessibility indicators, travel time to the closest market, cumulative accessibility, and entropy, we computed pair‑wise Spearman rank correlations (population‑weighted medians) at both country and pixel scales.
In Fig. \ref{fig:metric_correlations_sensitivity} panel \textbf{a} we observe the correlation matrix of the population-weighed medians of the accessibility metrics at the country level. In panel \textbf{b}, we increase the resolution of the correlation analysis at the pixel level. In this case, firstly, we compute the rank correlation coefficient for every country. Secondly, we calculate the average of the correlation coefficients across all countries in Africa as well as the average of the p-values. Results still display significant correlations, although, expectedly, the magnitude of the correlation decreases significantly due to the increased resolution.

In Fig. \ref{fig:metric_correlations_sensitivity}, panels \textbf{c} and \textbf{d} show the sensitivity analysis of the number of markets within a given travel time (cumulative accessibility), changing the threshold radius for Ethiopia. 
More specifically, in \textbf{c} we display the population average with the standard deviation and the maximum value, and in \textbf{d} we portray the relative variation (similar to a derivative, see eq. \ref{eq:deriv}) of the population average. We chose to study the sensitivity of this metric in Ethiopia because it is a country with enough data, both from OSM and the WFP, to perform an appropriate analysis.

\begin{equation}\label{eq:deriv}
    y'_i = \frac{y(x_{i+1}) - y(x_i)}{x_{i+1}-x_i}
\end{equation}

In eq. \ref{eq:deriv}, $x$ is the travel time threshold, $y$ is the population average and $i$ is the index of the time threshold sequence.

The results in Fig. \ref{fig:metric_correlations_sensitivity} \textbf{c} and \textbf{d} support the robustness of the proposed 30-minute motorized travel time threshold for the cumulative accessibility.
Varying the travel‑time threshold from 5 to 180 min in Ethiopia, mean accessible‑market count remains stable around $x$=30min, and the derivative of the mean approaches zero in the 25–40min window, validating our chosen threshold.

\section{Population divided in rural-urban classes within UN Africa Subregion}

Table \ref{tab:populations} disaggregates the population obtained by Worldpop in 2020 by urbanisation class (Urban, Suburban, Rural, Hinterland) for each of the five U.N. African sub‑regions. Percentages highlight how Western and Eastern Africa are dominated by suburban settlements, whereas Middle Africa retains the largest hinterland share.
 
\begin{table}[ht!]
\caption{\textbf{Population in urban-rural areas for UN Africa Subregions}: Population in millions and percentage in the urban, suburban, rural and hinterland classes in the five UN Africa Subregions.}
\label{tab:populations}
\begin{tabular}{@{}cccccc@{}}
\toprule
\multirow{2}{*}{UN Subregion} & \multicolumn{5}{c}{Population (millions) (population share)}             \\
& Urban          & Suburban       & Rural         & Hinterland    & Total  \\ \midrule
Northern Africa              & 113.38 (46 \%) & 98.97 (40 \%)  & 14.20 (6 \%)  & 19.32 (8 \%)  & 245.87 \\
Western Africa                & 105.42 (26 \%) & 223.25 (56 \%) & 54.77 (14 \%) & 17.78 (4 \%)  & 401.21 \\
Middle Africa                 & 40.72 (23 \%)  & 72.38 (40 \%)  & 35.09 (20 \%) & 31.18 (17 \%) & 179.37 \\
Eastern Africa                & 53.26 (15 \%)  & 201.73 (56 \%) & 66.65 (19 \%) & 38.44 (11 \%) & 360.08 \\
Southern Africa               & 47.42 (31 \%)  & 60.85 (40 \%)  & 32.86 (22 \%) & 9.91 (7\%)    & 151.04 \\ \bottomrule
\end{tabular}
\end{table}

To improve transparency, we compare our aggregated URCA classes and United Nations (UN) Population Division urbanization estimates in Table \ref{tab:UN_diff_rururb}. This comparison shows that our “Suburban” class contains a larger proportion of the population compared to the UN “Towns” category (~16\% more on average), whereas our “Urban” class captures a smaller share relative to the UN “Cities” category (~16\% less on average). This suggests that the main difference arises from how peri-urban and connected settlements are classified, rather than from a systematic underestimation of rural populations. 

Importantly, the purpose of this classification is not to reproduce official demographic statistics but to stratify populations by functional spatial accessibility to urban centers. The main conclusions remain robust under this interpretation: accessibility declines consistently along the continuum from urban and connected regions toward increasingly remote areas.

\begin{table}[ht!]
    \centering
    \caption{Share of the population of each rural-urban class for the African UN subregions according to the United Nations' official estimates, and difference in share between the URCA 4-level division and the UN categories. The comparison uses the following mapping: cities = urban, towns = suburban, and rural UN = rural + hinterland. Positive differences indicate that the URCA class has a larger population share than the corresponding UN category, and negative values indicate a smaller share.}
    \label{tab:UN_diff_rururb}
    \begin{tabular}{lrrrrrr}
        \toprule
        \multirow{2}{*}{Subreg.} 
        & \multicolumn{3}{c}{UN Division (\%)} 
        & \multicolumn{3}{c}{Difference (URCA -- UN) (\%)} \\
        \cmidrule(lr){2-4} \cmidrule(lr){5-7}
        & Cities & Towns & Rural 
        & Urban--Cities & Suburb.--Towns & Rural+Hint.--Rural UN \\
        \midrule
        NA & 57   & 29   & 14   & -11   & 11   & 0 \\
        WA & 43.6 & 29.3 & 27.1 & -17.6 & 26.7 & -9.1 \\
        MA & 51.5 & 21.4 & 27.1 & -28.5 & 18.6 & 9.9 \\
        EA & 25.7 & 41.2 & 33.1 & -10.7 & 14.8 & -3.1 \\
        SA & 46.5 & 29.2 & 24.4 & -15.5 & 10.8 & 4.6 \\
        \midrule
        Average & -- & -- & -- & -16.66 & 16.38 & 0.46 \\
        \bottomrule
    \end{tabular}
\end{table}

\newpage
\section{Summary of accessibility metrics}

\begin{table}[ht!]
\caption{\textbf{Pros and Cons of market accessibility metrics}: Schematic comparison of the advantages and disadvantages of the three accessibility metrics used in this work.}
\label{tab:pros_cons}
\begin{tabular}{@{}clll@{}}
\cmidrule(l){2-4}
 &
  \multicolumn{1}{c}{\textbf{CUMULATIVE}} &
  \multicolumn{1}{c}{\textbf{TRAVEL TIME}} &
  \multicolumn{1}{c}{\textbf{ENTROPY}} \\ \midrule
\rowcolor[HTML]{FFFFFF} 
Concept &
  \multicolumn{1}{c}{\cellcolor[HTML]{FFFFFF}Number of markets} &
  \multicolumn{1}{c}{\cellcolor[HTML]{FFFFFF}Closeness} &
  \multicolumn{1}{c}{\cellcolor[HTML]{FFFFFF}Spatial Distibution} \\ \midrule
\rowcolor[HTML]{FFFFFF} 
Pros &
  \begin{tabular}[c]{@{}l@{}}1. Interpretable units\\ 2. Considers quantity\end{tabular} &
  \begin{tabular}[c]{@{}l@{}}1. Interpretable units\\ 2. No parametric dependence\end{tabular} &
  \begin{tabular}[c]{@{}l@{}}1. Smaller value range\\ 2. Bridges cumulative \\ \quad and travel time\end{tabular} \\ \midrule
\rowcolor[HTML]{FFFFFF} 
Cons &
  \begin{tabular}[c]{@{}l@{}}1. Bias towards \\ \quad populated areas\\ 2. Arbitrary\\ \quad threshold radius\end{tabular} &
  \begin{tabular}[c]{@{}l@{}}1. Does not consider quantity\\ 2. Assumes consumers only \\
  \quad  travel to the closest market\end{tabular} &
  \begin{tabular}[c]{@{}l@{}}1. Not interpretable units\\ 2. Assumed travel behaviour \\ \quad in the absence of data\\ 3. Lack of references \\ \quad in the literature\end{tabular} \\ \bottomrule
\end{tabular}%
\end{table}

\section{OSM and WFP data coverage in Ethiopia and Nigeria}
\label{app:osm_wfp_examples}

\begin{figure}[ht!]
    \centering
    \begin{tabular}{ll}
        \textbf{a} & \textbf{b} \\
    \includegraphics[width = 0.49\linewidth]{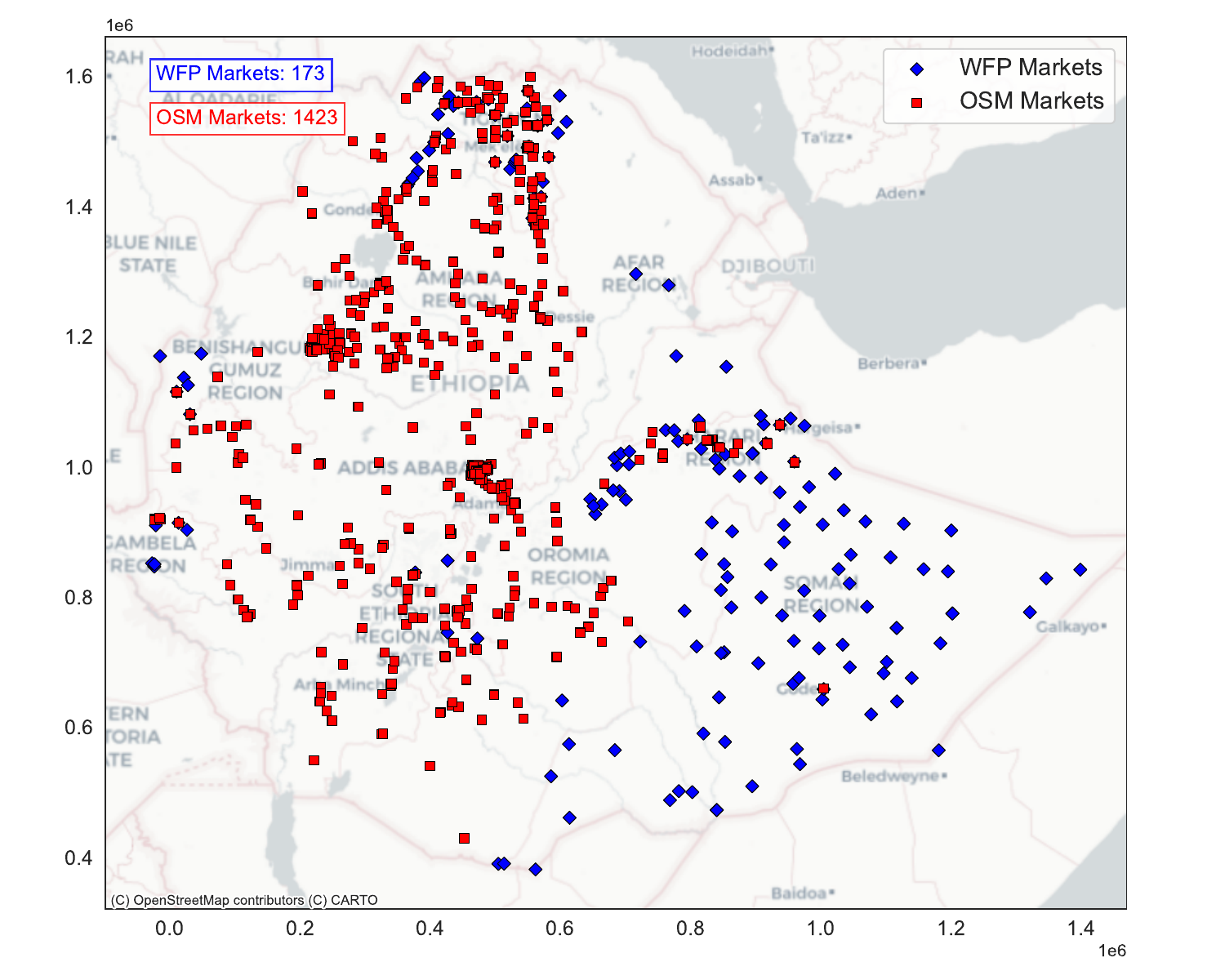} &
    \includegraphics[width = 0.49\linewidth]{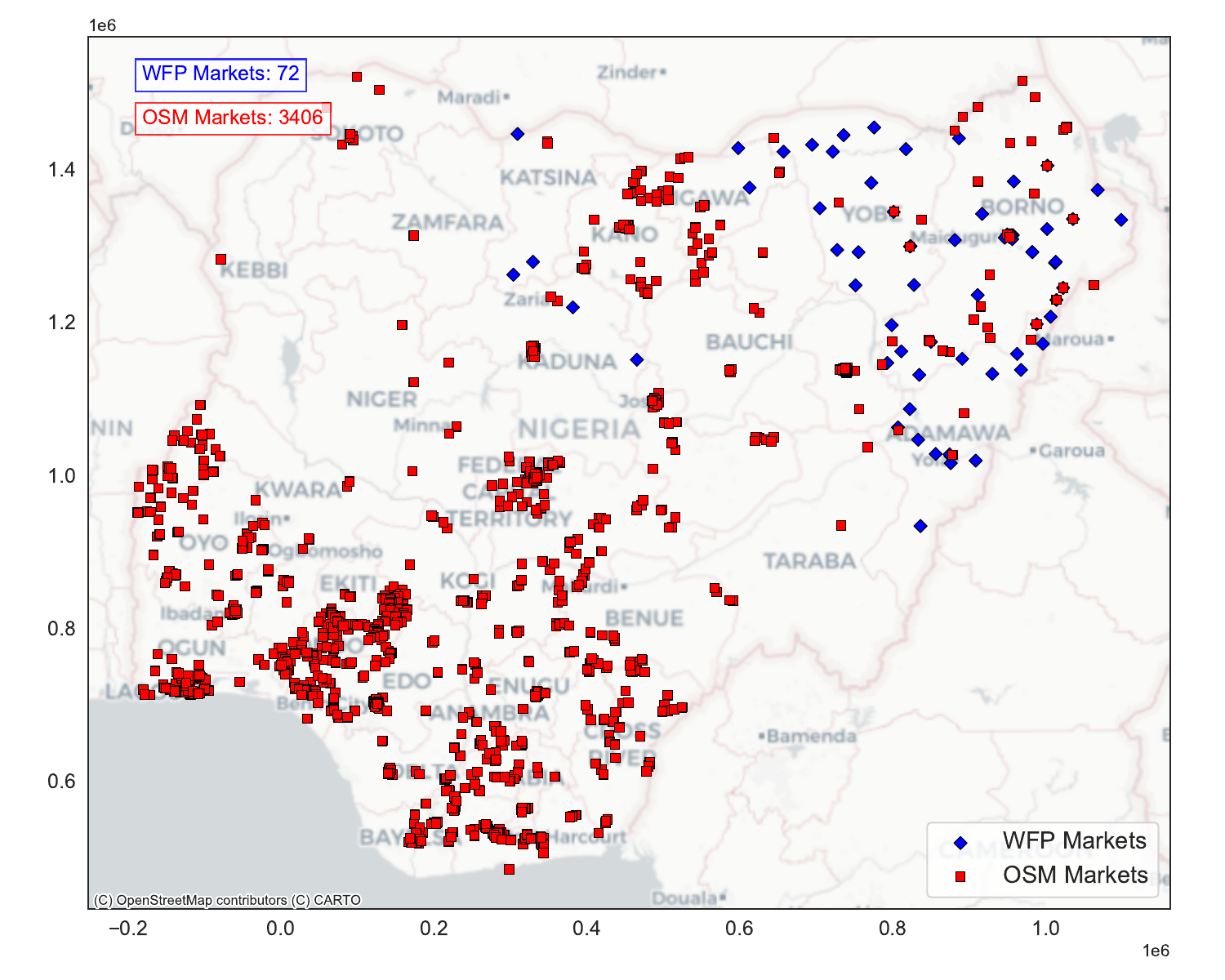}
    \end{tabular}

    \caption{Food shops and markets collected from OSM and WFP in Ethiopia (\textbf{a}) and Nigeria (\textbf{b}). These correspond to the raw data before the 15-minute clustering process.}
    \label{fig:points} 
\end{figure}

To visualise complementarity between data sources, In Fig. \ref{fig:points} maps raw OpenStreetMap (OSM) amenities and WFP market points in Ethiopia and Nigeria before centroid clustering. WFP points concentrate in food‑insecure or sparsely mapped regions (e.g., Somali Region in Ethiopia, north‑east Nigeria), underscoring the value of merging the two datasets.

\begin{figure}[ht!]
    \centering
    \includegraphics[width =\linewidth]{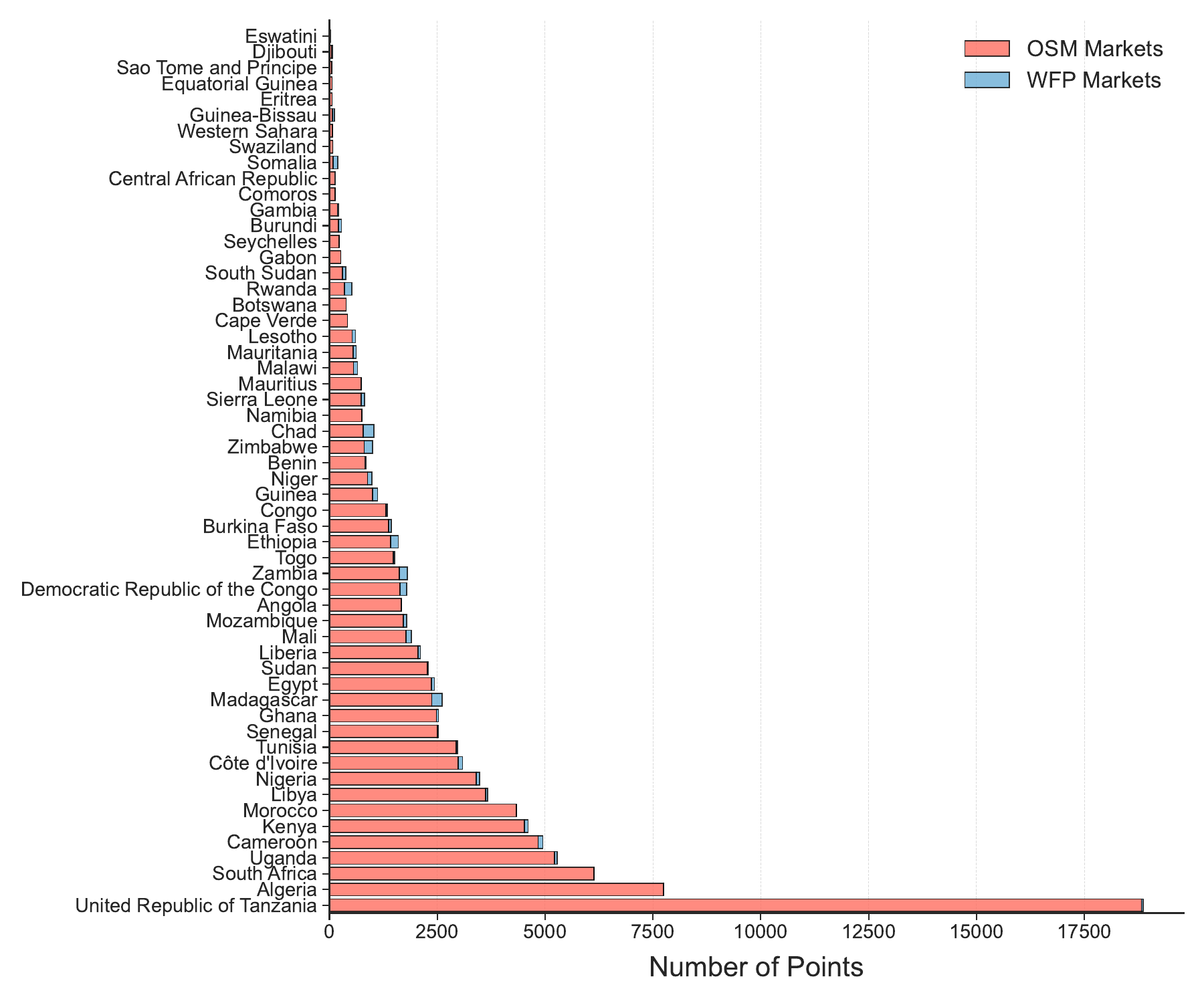}

    \caption{Number of OSM (red) and WFP (blue) markets per country. We use a logarithmic y-axis for an improved visualization of the countries with WFP markets.}
    \label{fig:MFI} 
\end{figure}

Figure \ref{fig:MFI} ranks all 54 African countries by the number of raw market points from each source (log‑scaled axis). Disparities highlight where targeted data‑improvement campaigns could most benefit accessibility estimates.

\section{Extracting market centroids}\label{app:centroids}

\begin{figure}[ht!]
    \centering
    \begin{tabular}{ll}
        \textbf{a} &  \textbf{b}\\
        \includegraphics[width = 0.49\linewidth]{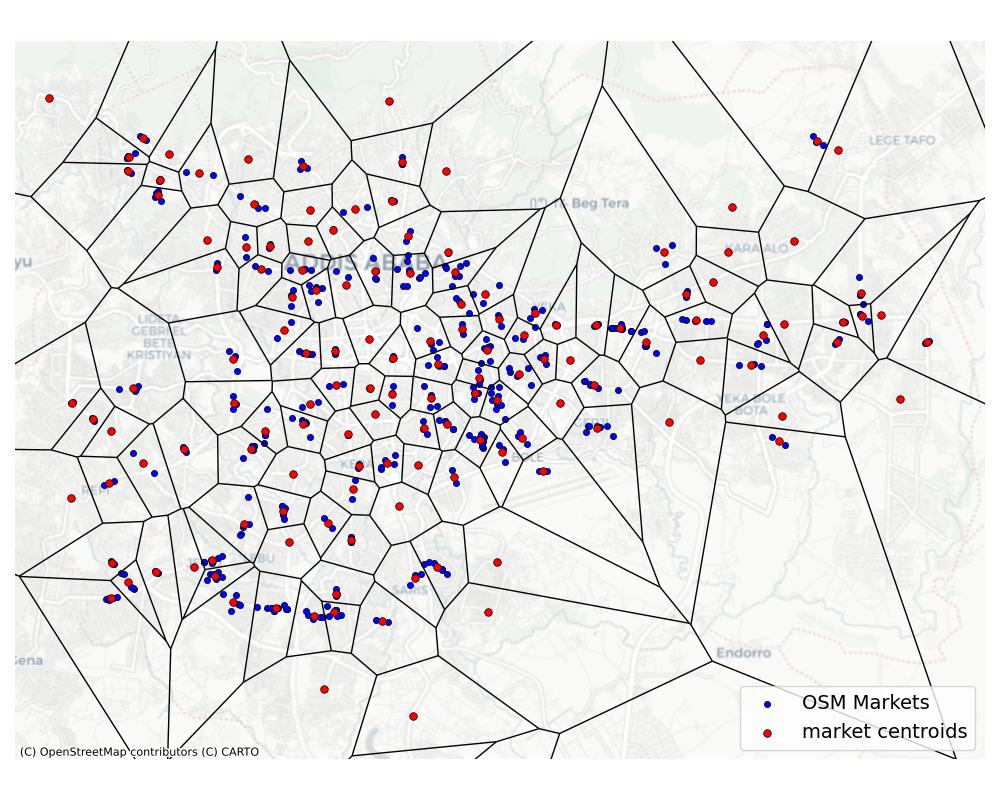}&
        \includegraphics[width = 0.49\linewidth]{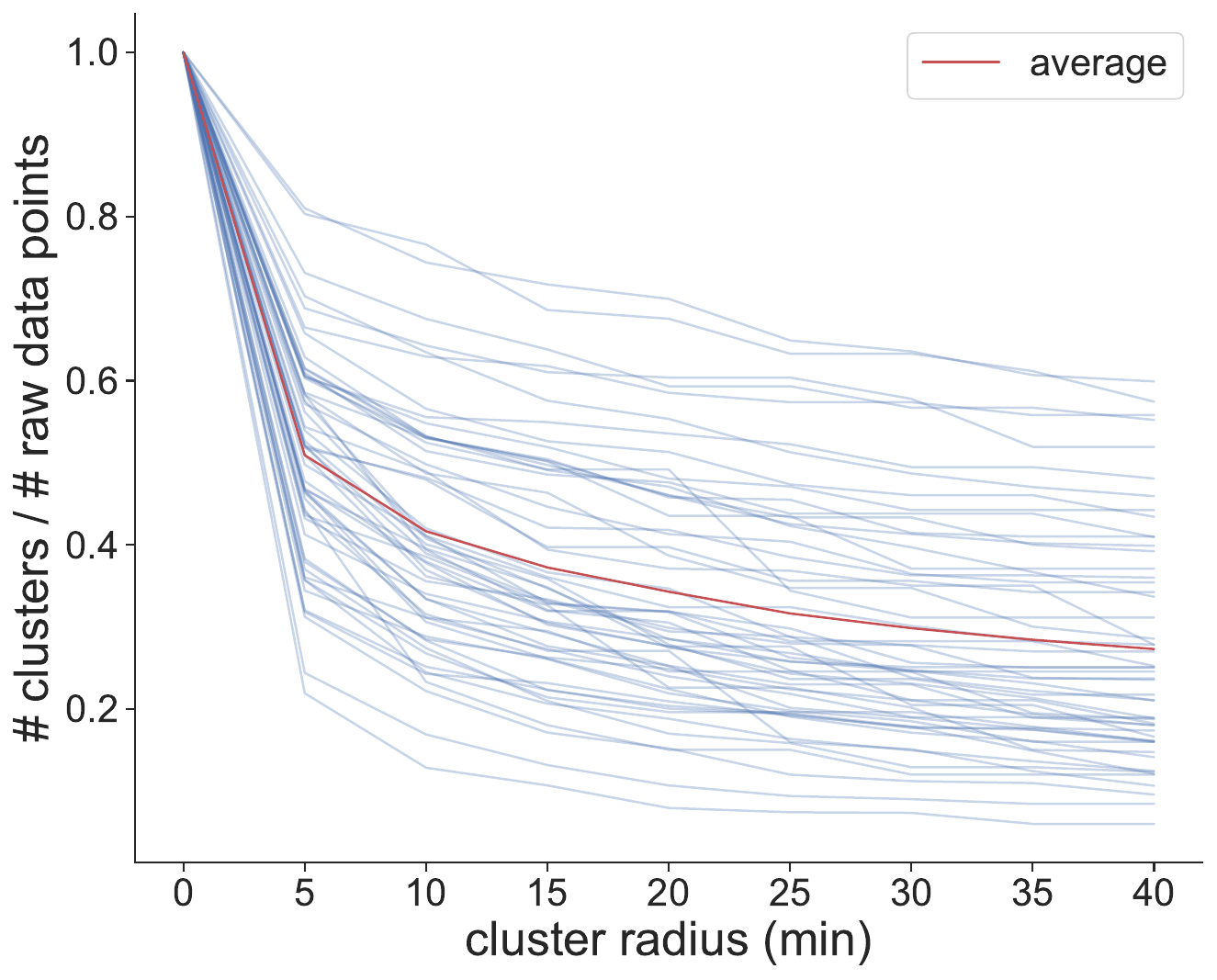}
    \end{tabular}
    \caption{\textbf{Clustering example and sensitivity analysis} (\textbf{a}) 15-minute cluster centroids of the OSM food shops with the respective Voronoi cells in Addis Abeba. Sensitivity analysis (\textbf{b}) of the number of clusters produced for different threshold radius for all African countries}
    \label{fig:clusters} 
\end{figure}

We transform raw points into market centroids by agglomerating amenities within a 15‑min walking radius. Fig. \ref{fig:clusters} \textbf{a} shows the resulting centroids and Voronoi catchments for Addis Ababa. Figure \ref{fig:clusters} \textbf{b} plots the ratio nclusters/npoints versus candidate radii across Africa. Initially, for small cluster radius, the relative number of clusters drops significantly. The curve shows a slower decrease for 15 minutes and beyond, indicating that small changes in cluster radius do not significantly impact the resulting market distribution overall.

\section{Class distribution of IPC data}
\begin{figure}
    \centering
    \includegraphics[width=\linewidth]{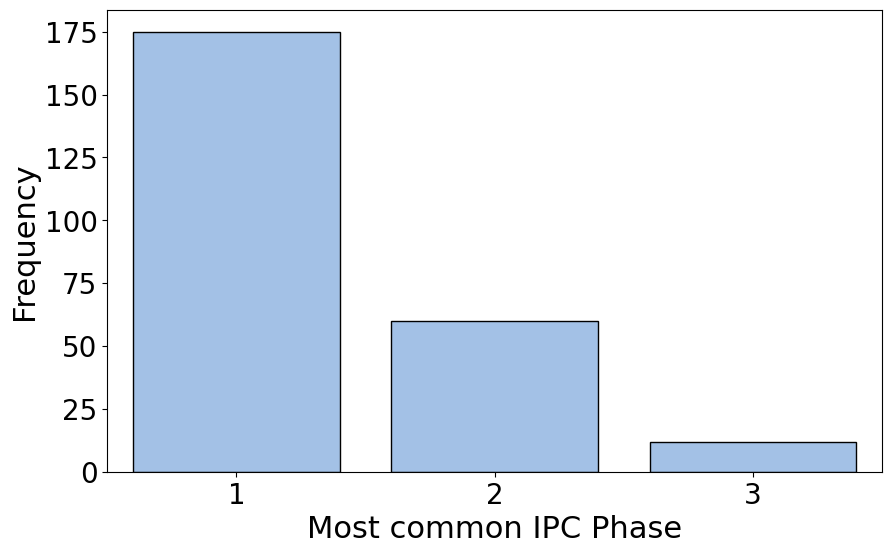}
    \caption{Distribution of the most prevalent IPC phase by population across administrative level-1 regions in Africa, showing the number of regions in each phase.}
    \label{fig:IPC_barplot}
\end{figure}

Figure \ref{fig:IPC_barplot} shows the class imbalance in IPC phases, with IPC phase 1 being the most prevalent across all Africa.

\section{External comparison of friction-surface travel-time estimates}

To assess the robustness of the friction-surface-based travel-time estimates used in this study, we conducted an external comparison against travel times computed using a road-network-based routing engine. Specifically, we used the OpenRouteService (ORS) API, which calculates shortest-path travel times on the OpenStreetMap (OSM) road network.

\subsection{Sampling procedure}

We randomly sampled 100 origin–destination (OD) pairs across the study area. Origins correspond to randomly selected raster pixels from the 30 arc-second friction surface, while destinations correspond to randomly selected market locations derived from the harmonized OSM–WFP dataset.
Because network-based routing requires mapped road connectivity, we restricted the sampled OD pairs to cases where both the origin and the destination were located within 2 km of a mapped road segment in OSM. This restriction ensures that ORS can compute a valid route. It should be noted that this constraint excludes some remote areas lacking mapped road infrastructure, where friction surfaces still allow travel-time estimation via terrain-based routing.
For motorized travel, we computed travel times using ORS driving mode. For walking travel, we used the ORS walking profile. To facilitate interpretation, we report results both for the full set of sampled OD pairs and for subsets restricted by Euclidean distance thresholds ($\leq$ 80 km for motorized travel; $\leq$ 20 km for walking), as extremely long-distance walking routes may produce unstable comparisons due to modeling differences.

\subsection{Comparison methodology}

\begin{figure}[ht!]
    \centering
    \begin{tabular}{ll}
        \textbf{a} & \textbf{b} \\
        \includegraphics[width=0.48\linewidth]{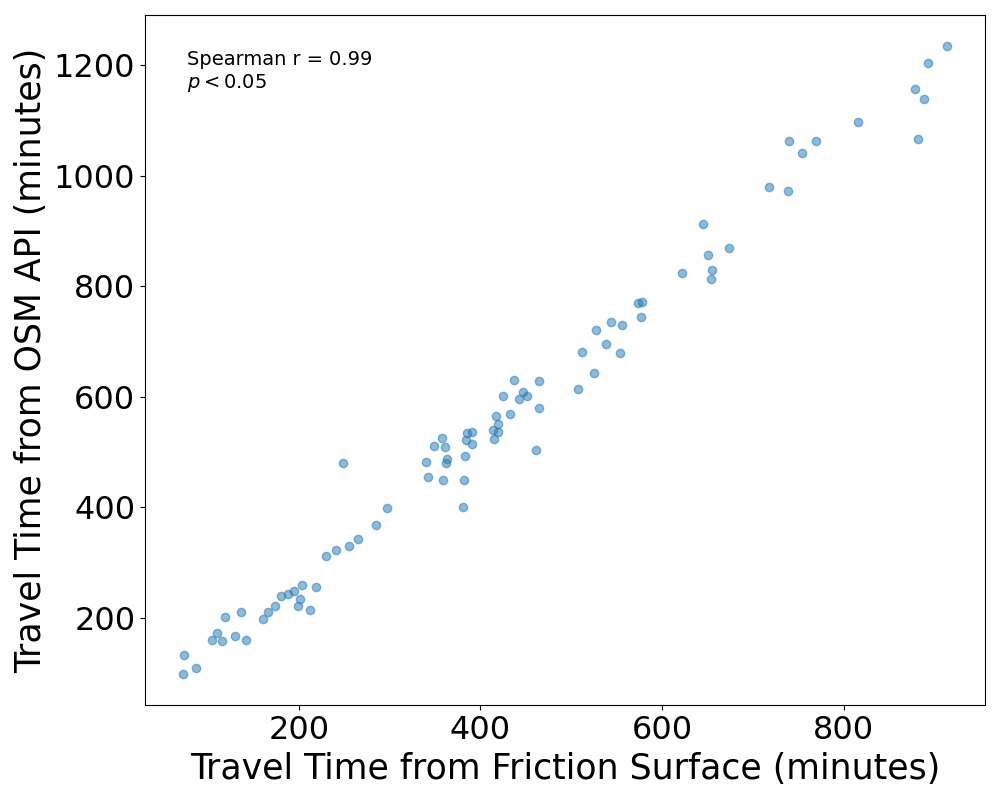} & \includegraphics[width=0.48\linewidth]{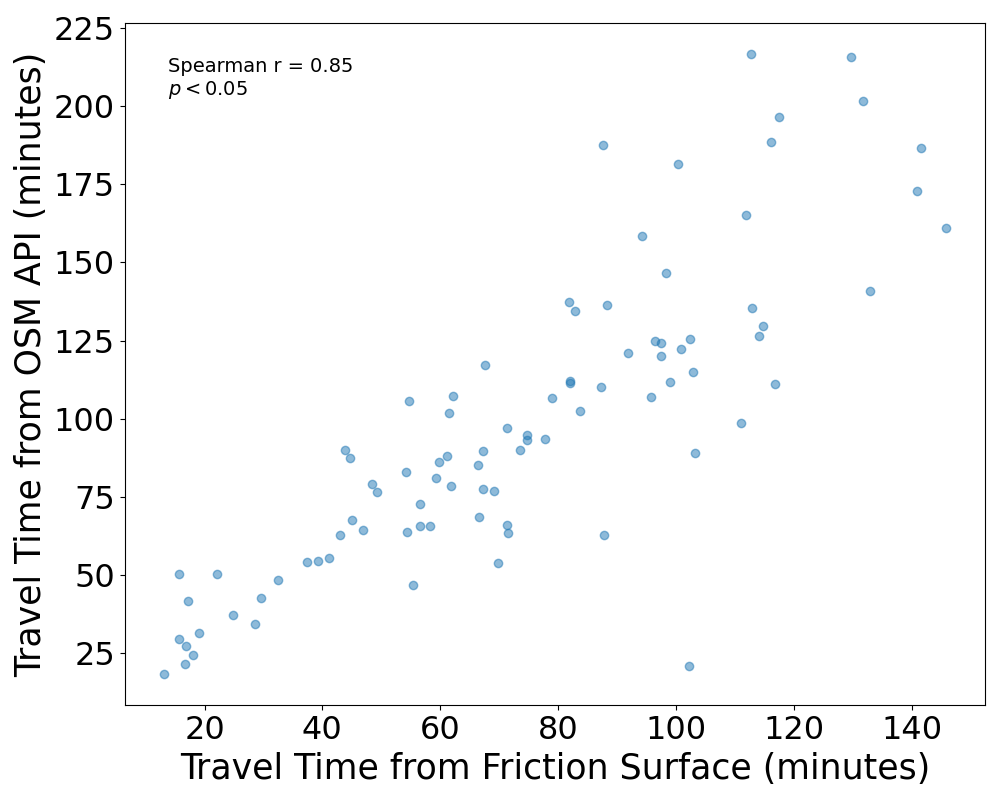} 
    \end{tabular}
    
    \caption{Comparison between motorized travel times computed using the friction-surface approach (this study) and travel times obtained via OpenRouteService routing on the OpenStreetMap road network, based on 100 samples of origin–destination pairs. (\textbf{a}) Subset of OD pairs with Euclidean distance $\leq$ 80 km. (\textbf{b}) Full set of sampled OD pairs. Differences between estimates reflect the distinct modelling assumptions of continuous friction-surface routing versus road-network-constrained routing and potential incompleteness of the mapped road network.}
    \label{fig:tt_val_moto}
\end{figure}

\begin{figure}[ht!]
    \centering
        \includegraphics[width=0.5\linewidth]{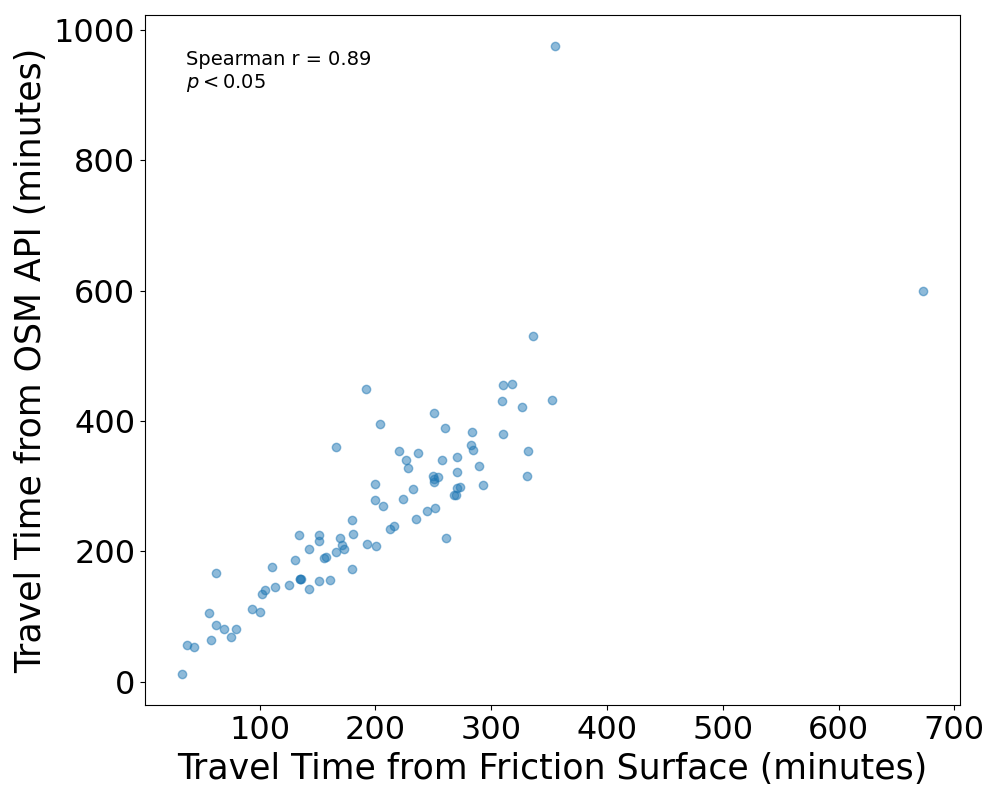} 
    \caption{Comparison between walking travel times computed using the friction-surface approach (this study) and travel times obtained via OpenRouteService routing on the OpenStreetMap pedestrian/road network, based on randomly sampled origin–destination pairs. Origins correspond to raster pixels and destinations correspond to randomly selected market locations. OD pairs were restricted to cases where both endpoints were within 2 km of a mapped road to ensure network routability. For interpretability, results are shown for OD pairs with Euclidean distance $\leq$ 20 km. Differences between estimates reflect the distinct modeling assumptions and potential incompleteness of mapped pedestrian infrastructure.}
    \label{fig:tt_val_walk}
\end{figure}

For each OD pair, we computed:
\begin{itemize}
    \item Travel time using the friction-surface approach described in the Methods (MCP\_Geometric algorithm applied to motorized and walking friction rasters).
    \item Travel time using OpenRouteService routing on the OSM road network.
\end{itemize}

We then compared the two sets of estimates using scatterplots and correlation analysis. The results (Supplementary Figures X–Y) show a consistent positive association between friction-surface and network-based travel times across both motorized and walking modes.
Differences between the two approaches are expected and reflect their distinct modeling assumptions:

\begin{enumerate}
    \item Friction-surface routing represents a continuous cost landscape incorporating roads, land cover, slope, and off-road travel, allowing movement across all terrain types.
    
    \item Network-based routing (ORS) is constrained to mapped road and pedestrian networks and may underestimate connectivity in areas where road mapping is incomplete.
\end{enumerate}

Spearman correlation coefficients were $\rho = 0.99$ for unbounded pairs and $\rho = 0.85$ for a maximum OD separation of 80 km for motorized transport. For walking, $\rho = 0.85$ with a maximum separation between the origin and the market of 20 km. These results indicate substantial agreement between the two estimation approaches.
Overall, the comparison indicates that while absolute travel-time estimates differ in some cases, the relative ranking of OD travel times remains broadly consistent across methods. This supports the robustness of the friction-surface approach for capturing large-scale spatial gradients in accessibility.